\documentclass[preprints,article,accept,pdftex,moreauthors]{mdpi}

\firstpage{1} 
\pubvolume{1}
\issuenum{1}
\articlenumber{0}
\pubyear{2024}
\copyrightyear{2024}
\datereceived{ } 
\daterevised{ } 
\dateaccepted{ } 
\datepublished{ } 
\hreflink{https://doi.org/} % If needed use \linebreak

\usepackage{siunitx}

\Title{Radiation Impedance of Rectangular CMUTs} 

\TitleCitation{Radiation Impedance of Rectangular CMUTs}

\Author{Shayan Khorassany \orcidA{}, Eric B. Dew\orcidB{}, Mohammad Rahim Sobhani and Roger J. Zemp \orcidC{}*}

\AuthorNames{Shayan Khorassany, Eric Dew, Mohammad Rahim Sobhani and Roger Zemp}

\AuthorCitation{Khorassany, S; Dew, E.; Sobhani, MR; Zemp, R.}

\address[1]{%
Department of Electrical and Computer Engineering, University of Alberta, Edmonton, 
Alberta, Canada T6G 1H9}

\corres{Correspondence: rzemp@ualberta.ca}

\abstract{
Recently, capacitive micromachined ultrasound transducers (CMUTs) with long rectangular membranes have demonstrated performance advantages over conventional piezoelectric transducers; however, modeling these CMUT geometries has been limited to computationally burdensome numerical methods. Improved fast modeling methods such as equivalent circuit models could help achieve designs with even better performance. The primary obstacle in developing such methods is the lack of tractable methods for computing the radiation impedance of clamped rectangular radiators. This paper presents a method which approximates the velocity profile using a polynomial shape model to rapidly and accurately estimate radiation impedance. The validity of the approximate velocity profile and corresponding radiation impedance calculation was assessed using finite element simulations for a variety of membrane aspect ratios and bias voltages. Our method was evaluated for rectangular radiators with width:length ratios from 1:1 up to 1:25. At all aspect ratios, the radiation resistance was closely modeled. However, when calculating the radiation reactance, our initial approach was only accurate for low aspect ratios. This motivated us to consider an alternative shape model for high aspect ratios, which was more accurate when compared with FEM. To facilitate development of future rectangular CMUTs, we provide a MATLAB script which quickly calculates radiation impedance using both methods.
}


\begin{document}

%%%%%%%%%%%%%%%%%%%%%%%%%%%%%%%%%%%%%%%%%%

\section{Introduction}
Over the last three decades, capacitive micro-machined ultrasound transducers (CMUTs) have demonstrated significant potential to replace piezoelectric-based ultrasound transducers. For ultrasound imaging arrays, the advantages of CMUTs include wide bandwidth, sensitive receive operation, and easier integration with electronics \cite{haller1996, Herickhoff2023, Joseph2022}. Traditionally, a CMUT linear array element is designed with many small CMUT cells operating in parallel. In this approach, each CMUT cell is a single vibrating membrane, and the behavior of the overall element is determined by the combined behavior of these oscillating membranes. These individual cells are typically designed with circular membranes; this is partially due to circular membranes being simple to model and predict owing to axial symmetry. Such circular membranes have been studied extensively \cite{first_CMUT, LaMura2022, Aditi2022}, although other works have also explored square \cite{maadi_square}, hexagonal \cite{Mukhiya2015, Ergun2005}, and rectangular membranes \cite{Wong2008}.

Although this multiple-membrane approach is straightforward, there are also some key drawbacks to this paradigm which limit the overall performance of CMUT elements. Due to the nonlinear nature of CMUT operation, CMUT membranes collapse when the DC bias voltage is increased beyond a certain threshold, termed the collapse voltage or pull-in voltage. In a CMUT element with a large number of membranes, fabrication nonuniformities lead to variations in collapse voltage between individual cells. To avoid operating in a regime with some collapsed and some uncollapsed membranes, CMUTs are typically biased at 80-90\% of the mean collapse voltage, where electromechanical efficiency is considerably reduced \cite{Yaralioglu2003}. Another key drawback is the potential for asynchronous membrane motion and mutual acoustic coupling effects, which further reduce CMUT performance \cite{Atalar2014, Maadi2016, Meynier2010}. As a result of these deleterious effects, it is common for CMUTs to have lower electromechanical efficiency and reduced transmit pressure compared to piezoelectric transducers, which has limited their adoption in ultrasound systems \cite{Herickhoff2023}.

To avoid these drawbacks, Dew \emph{et al.} recently proposed an approach to designing CMUTs with a single rectangular membrane per element as opposed to multiple membranes \cite{ericdew}. Such an approach demonstrated significantly improved electromechanical efficiency, and greatly improved transmit efficiency, outperforming not only other CMUT designs but also many piezoelectric transducers by up to 3-fold. However, the rectangular membranes in these devices had a width of $\lambda/8$ which is much smaller than the typical half-lambda pitch needed for a phased array \cite{ericdew}. The dimensions and performance of these single-membrane CMUTs could likely be further optimized if a robust modeling approach were available. 

Although computational methods such as finite element method (FEM) simulations are very powerful, it is desirable to develop analytical models which provide theoretical insight into the impacts of design parameters and faster runtime. One powerful analytical tool for modeling electro-mechanical devices is lumped equivalent circuit modeling. Equivalent circuit models are commonly used to predict CMUT behavior, owing to their computational advantages, and potential for integrated simulations with back-end electronics \cite{improved_koymen, Merrien2022}. Circuit models are commonly used for circular and square membrane geometries; however, it has been difficult to develop analogous models for rectangular membranes. Although rectangular geometries generally suffer from fewer symmetries, the primary obstacle in developing these models is the radiation impedance of the membrane.

Radiation impedance is a lumped parameter which describes the coupling between the acoustic medium and vibrating membrane. This quantity is a key parameter in circuit models for circular and square membrane geometries \cite{improved_koymen, koymen, Caronti2002, Merrien2022, maadi_square}; however, developing analogous expressions for rectangular membranes is challenging due to a lack of symmetry. The CMUT membrane is best described as a clamped radiator \cite{koymen}; however, a mathematically simpler approximation is to instead model the membrane as a radiating piston. Unfortunately, the rectangular geometry is sufficiently complicated that even in the case of piston motion, a closed-form solution has not been obtained \cite{TimMellow}.

Many previous approaches to analyzing clamped radiators have focused on modal analysis of rectangular plates. This approach is useful for studying the natural vibrations of these structures, but is more challenging to apply to a CMUT circuit model context. Moreover, many of these works focus on relatively low length:width aspect ratios, and are not validated with FEM simulations \mbox{\cite{yoo2010study, Li2001, beranek1971noise, Zhang2023, simple_calc}}. Given the high aspect ratios of single-membrane rectangular CMUTs, and added complexity of CMUT operation, there is an unmet need for further research into this topic.

In this paper, we present a method to calculate the radiation impedance of rectangular CMUTs. By using polynomial shape models to approximate the velocity profile of the fundamental mode, we were able to derive integrals which could be evaluated to calculate the radiation impedance. For exceptionally high aspect ratios, we repeated this method using a “1D” approximate velocity profile, which offered better agreement. These integrals were evaluated numerically using a MATLAB script which calculates the real and imaginary portions of the radiation impedance at a given frequency and aspect ratio. To validate our approach, we performed FEM simulations in COMSOL Multiphysics, which were used to assess to accuracy of our approximate velocity profile and the corresponding radiation impedance results. Our approach has close agreement with FEM for radiation resistance, and reasonable agreement for radiation reactance. Moreover, our MATLAB script can be run up to 4 orders of magnitude faster than the corresponding FEM simulations. This approach can be used to accelerate the development of rectangular CMUTs through rapid iterations over large parameter spaces in an equivalent circuit model. In fact, our method is used in a study by Dew \emph{et al.} in \mbox{\cite{eric2024}}. We have included our MATLAB script in the supplementary information section.

%%%%%%%%%%%%%%%%%%%%%%%%%%%%%%%%%%%%%%%%%%
\section{Background}
Radiation impedance [\unit{Rayl.m^{2}}] is a parameter that describes the transfer of energy between the transducer and the medium. This parameter is often critical in the design of acoustic radiators such as ultrasound transducers, as it impacts the resonance frequency and bandwidth of the device. Mathematically, the radiation impedance, $Z_R$, is defined as the ratio of total emitted power, $P_{tot}$ to the square of a reference velocity, $v_i$.

\begin{linenomath}
\begin{equation} \label{RI_def}
    Z_R = \frac{P_{tot}}{|v_i|^2}
\end{equation}    
\end{linenomath}

This reference velocity is usually selected as the spatial root-mean-square (RMS) average velocity of the membrane in accordance with the conventions used in circuit models \cite{improved_koymen}. Calculating radiation impedance requires calculating the total emitted power, which is evaluated using a surface integral as shown in \eqref{ZR}

\begin{linenomath}
\begin{equation} \label{ZR}
    Z_R = \frac{ \iint_S \tilde{p}(\textbf{r}) v^*_n(\textbf{r}) \,ds }{|v_i|^2}
\end{equation}    
\end{linenomath}

In this expression, $\tilde{p}(\textbf{r})$ represents the pressure, and $v_n(\textbf{r})$ is the normal component of particle velocity to the surface element located at $\textbf{r}$. These quantities are both expressed as phasors with the time-dependent portion omitted. The pressure phasor can be computed using the particle velocity and the boundary-dependent Green's function of the problem, as discussed further in Section \ref{derivation}. Therefore, the membrane's velocity profile is a key assumption in the computation of radiation impedance.

%%%%%%%%%%%%%%%%%%%%%%%%%%%%%%%%%%%%%%%%%%
\section{Methods}
\subsection{Velocity Profile}

When calculating radiation impedance, it is first necessary to obtain an accurate velocity profile. This is particularly challenging in the case of rectangular plates due to the lack of cylindrical symmetry and the potentially vague plate regime (membrane, thin, or thick plate). In this section, we obtain an approximate expression for the velocity profile of a clamped rectangular plate, then validate our approach using FEM simulations of rectangular CMUTs with dimensions similar to previously fabricated devices \mbox{\cite{ericdew, ericdew2}}. 

Although we use the terms "membrane" and "thin plate" interchangeably in this paper, they have separate definitions in plate theory. It is common to model CMUT membranes according to classical plate theory and to use "thin plate" simplifying assumptions regarding negligible stress components. Thus, assuming an isotropic material, the static deflection of these devices can be described by \eqref{plate_eq} \cite{Funding2015, Wygant2008, Timoshenko1959, Ventsel2001txtbook}.

\begin{linenomath}
\begin{equation} \label{plate_eq}
     \frac{\partial^4 w}{\partial x^4} + 2\frac{\partial^4 w}{\partial x^2 \partial y^2} + \frac{\partial^4 w}{\partial y^4} = \frac{p}{D},
\end{equation}
\end{linenomath}
where $w(x,y)$ is the out-of-plane membrane deflection, $p$ represents the pressure on the membrane, and $D$ is the flexural rigidity of the plate. 

An exact solution to \eqref{plate_eq} for a uniform pressure load does not exist, but expansions utilizing trigonometric or polynomial basis functions have been used to approximate the profile \cite{Rahman2013, Olszacki2009, Taylor2004}. The expansion coefficients in these approaches are functions of the length:width aspect ratio of the rectangular plate, as well as its relative thickness. The devices we consider here have a thickness of \qty{4.9}{\micro\m}, a width of \qty{150}{\micro\m}, and lengths from \qty{150}{\micro\m} to \qty{3750}{\micro\m}.

To calculate the radiation impedance of a CMUT, an approximate velocity profile in the fundamental resonance mode while deflected by a DC bias is required. The non-uniform electrostatic load on the plate and its static deflection due to the DC bias make this problem much more involved compared to previous studies \mbox{\cite{Rahman2013, Olszacki2009, Taylor2004, ZHANG2010}}. Therefore, the rigorous analytical or numerical treatment of this problem is remarkably challenging.  

For the approximate velocity profile, we took inspiration from the first term of the polynomial-based deflection profiles used in \cite{Timoshenko1959, Blasquez1987, anis_plates, Olszacki2009}. This 4-th order polynomial expression satisfies appropriate boundary conditions for the velocity profile of a clamped plate. Thus, the membrane velocity $v(x,y)$ is given by \eqref{v(x,y)}.

\begin{figure}[H]
\includegraphics[width=10cm]{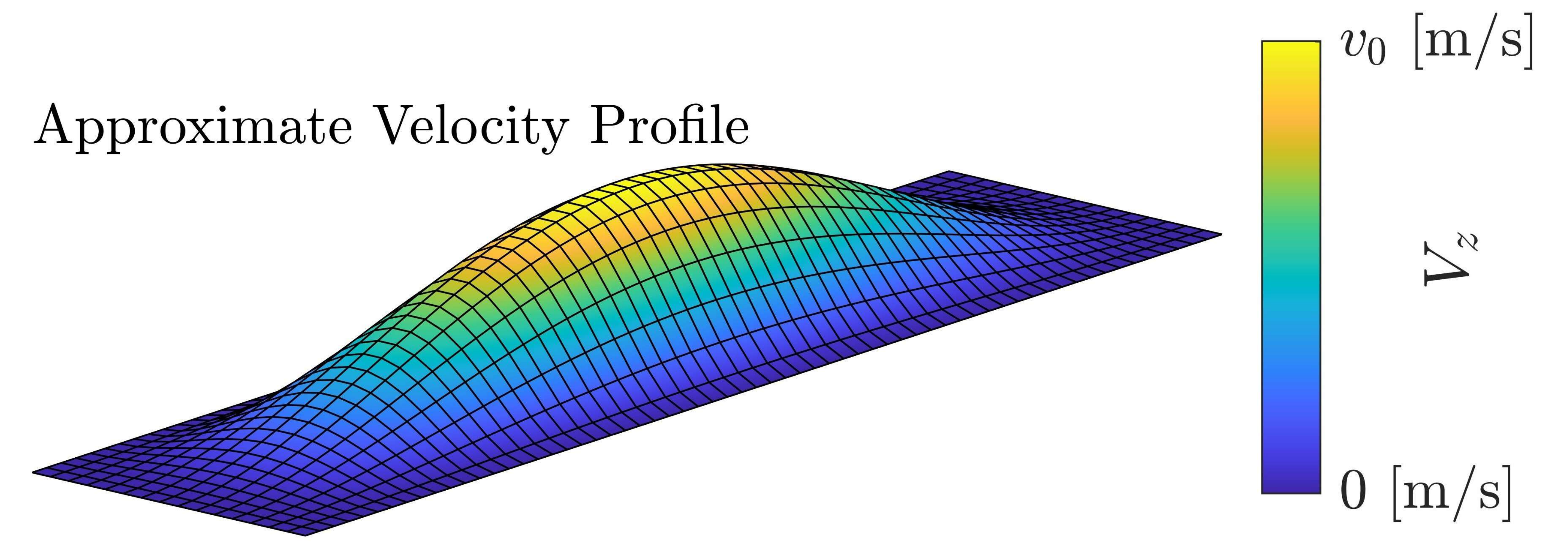}
\caption{The approximate velocity profile of a rectangular CMUT with an aspect ratio of 4 is plotted using the proposed expression of \eqref{v(x,y)}. \label{example_vel_prof}}
\end{figure}

\begin{linenomath}
\begin{equation}\label{v(x,y)}
    v(x,y) = v_0 \left( 1 - \left(\frac{x}{a}\right)^2 \right)^2 \left( 1 - \left(\frac{y}{b}\right)^2 \right)^2,
\end{equation}    
\end{linenomath}
where $v_0$ is the peak velocity of the plate. The half-width in the x direction is represented by $a$, while the half-length is represented by $b$. A special case of \eqref{v(x,y)} with the aspect ratio ($b/a$) of 4 is depicted in Figure~\ref{example_vel_prof}.

The 4-th order polynomial shape function in \eqref{v(x,y)} is usually complemented by additional higher-order terms to describe the static deflection of square membranes. However, such additional terms are omitted from our expression for two main reasons. The first reason is mathematical convenience, as calculations for the emitted power used to obtain radiation impedance are already computationally cumbersome, and it is desirable to obtain a tractable result. 

\begin{figure}[H]
\includegraphics[width=9cm]{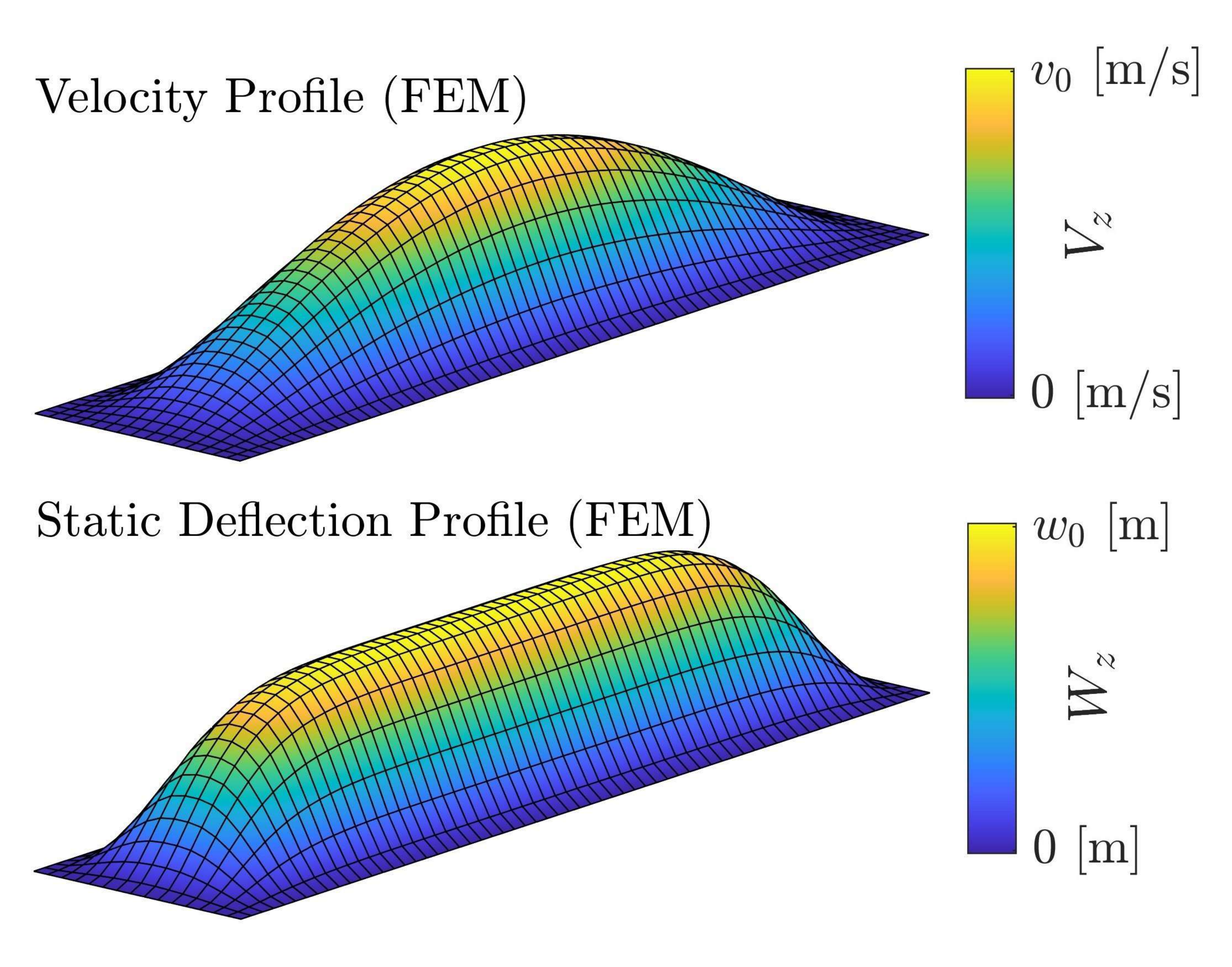}
\caption{The velocity profile of a rectangular 1:4 CMUT is depicted at the top while the static deflection is shown in the bottom half of the figure. For the velocity profile, the DC bias was set to \qty{20}{V} and the AC driving amplitude to \qty{1}{V}. The static deflection profile is depicted for the DC bias of \qty{50}{V}. These results were extracted from FEM simulations discussed in Section~\ref{FEMVal}. Note the significant difference between the profile shapes. \label{vel_vs_defl_prof}}
\end{figure}

Moreover, one key difference between rectangular membranes and other more typical membrane geometries is that the velocity profile may differ substantially from the static deflection profile \cite{Zhang2012, Tanaka2009}. The static deflection profile and the velocity profile of a 1:4 rectangular CMUT simulated by FEM are shown in Figure~\ref{vel_vs_defl_prof}. Thus, it is necessary to validate the accuracy of any proposed velocity profile using FEM simulations. An additional benefit of using FEM to validate our approach, is that FEM simulations account for higher order modes that are neglected in our approach. By comparing our result to FEM, the impact of these higher order modes can be assessed. 

\subsection{FEM Validation} \label{FEMVal}
\begin{figure}[H]
\begin{adjustwidth}{-\extralength}{0cm}
\centering
\includegraphics[width=12cm]{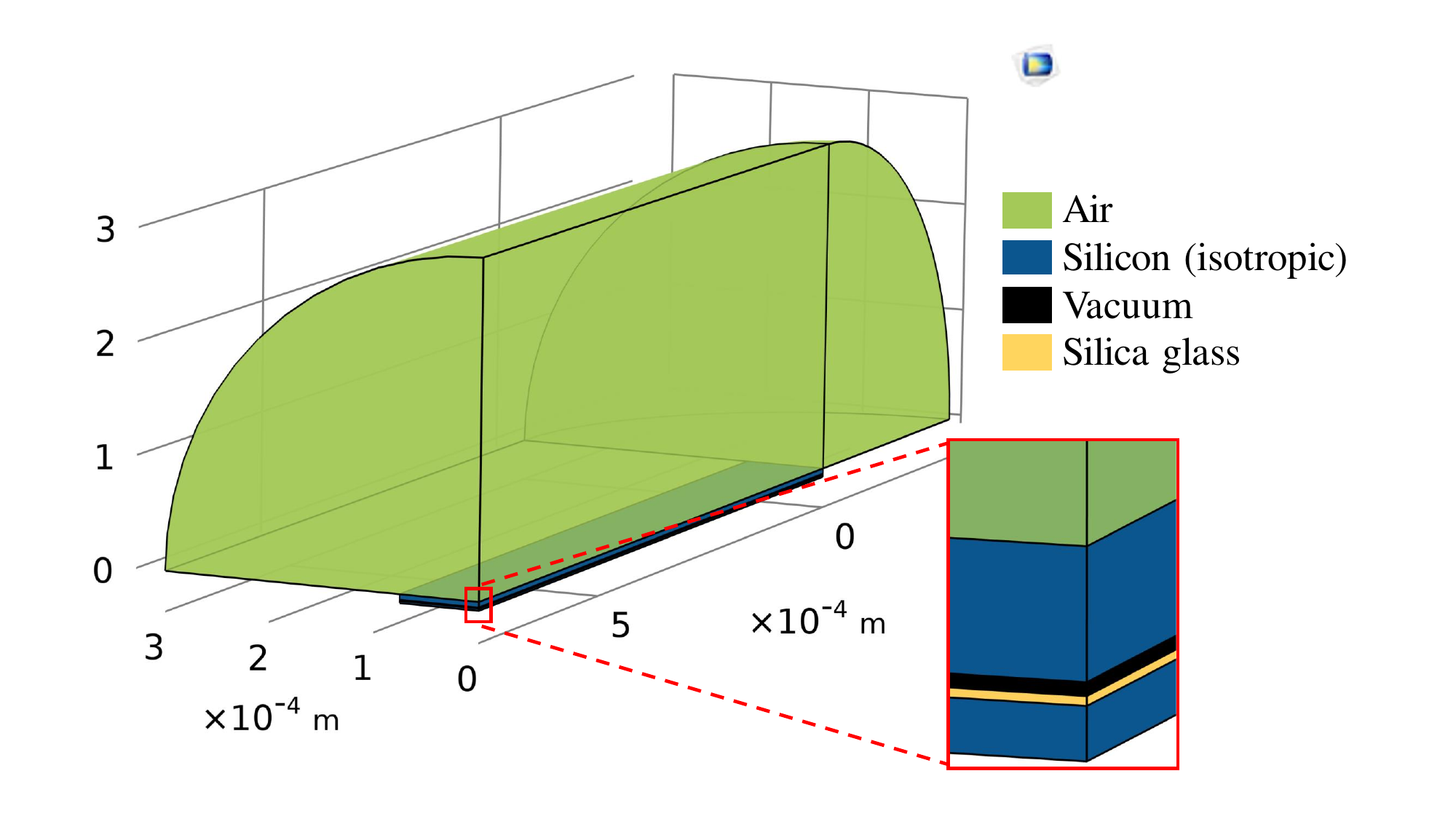}
\end{adjustwidth}
\caption{This figure shows the rectangular 1:10 CMUT simulation with COMSOL Multiphysics. A quarter of the space is simulated to reduce computational cost. The boxed corner is equivalent to the center of the CMUT. \label{diagramCOMSOL}}
\end{figure}

To investigate the validity of our approximate velocity profiles and the corresponding radiation impedance, 3D simulations of rectangular CMUTs were performed using COMSOL Multiphysics 6.2 (Comsol Inc., Burlington, MA). A diagram of the simulated geometrical cases (1:10 rectangular CMUT) is given in Figure~\ref{diagramCOMSOL}. These simulations were set up similarly to \cite{Mao_2017}. To confirm the accuracy of our FEM simulation method, we reproduced some of the results reported in \cite{Meynier2010} and \cite{maadi_square} using the boundary conditions and configurations described in this section.

To save computation time, a quarter of the membrane was simulated with the appropriate symmetry boundary conditions to obtain the same result as a full membrane. Across all these simulations, the width of the membrane was kept as \qty{150}{\micro\m} while the length differed for each case of aspect ratio, ranging from \qty{150}{\micro\m} to \qty{3750}{\micro\m}. The thickness of the membrane was kept as \qty{4.9}{\micro\m}. For the physics modules, the Pressure Acoustics Frequency Domain package was used alongside the Solid Mechanics and Electrostatics packages. The multiphysics interfaces included Acoustic-Structure Boundary and Electromechanical Forces. The silicon membrane was simulated using the parameters in the COMSOL materials library (Young's modulus: \qty{170}{GPa}, Poisson's ratio: \qty{0.28}).

The medium of propagation was air, modeled by Atmosphere attenuation in the Pressure Acoustics module. The speed of sound in this medium was \qty{343}{m/s} and the density was \qty{1.21}{kg/m^3}. The outer boundary of the medium was assigned the Perfectly Matched Boundary (PMB) condition. The dimensions of the acoustic medium and corresponding PMB were selected to considerably exceed the minimum requirements specified by COMSOL’s documentation. 

To ensure accuracy for every simulated frequency, the size of the medium was specified based on the lowest frequency, while the mesh size was selected based on the highest frequency. To accommodate the limitations of our hardware while maintaining accuracy, we performed separate simulations for the low and high frequencies. The accuracy of our approach was further verified by increasing the size of the medium and number of mesh elements in subsequent iterations. In these subsequent iterations, results varied by less than 0.1\%.

Through these simulations, the velocity profile of rectangular CMUTs of varying aspect ratios was obtained at different frequencies. This frequency sweep method proved to be much faster than eigenfrequency studies and helped us see the evolution in the velocity profile with changes in frequency. To ensure the fundamental mode of the membrane was accurately modeled, and that our frequency resolution was adequate, we used an iterative process with multiple simulations per aspect ratio. 

One important consideration is the bias voltage applied to the CMUT which has been shown to impact the velocity profile \cite{koymen}. To study this impact, we simulated each aspect ratio at 20\% and 90\% of the collapse voltage. The discrepancy between the FEM profile and our approximation was quantified using the absolute relative error (ARE) over the membrane. This metric is defined in \eqref{ARE}.

\begin{figure}[H]
\begin{adjustwidth}{-\extralength}{0cm}
\centering
\includegraphics[width=4.5cm]{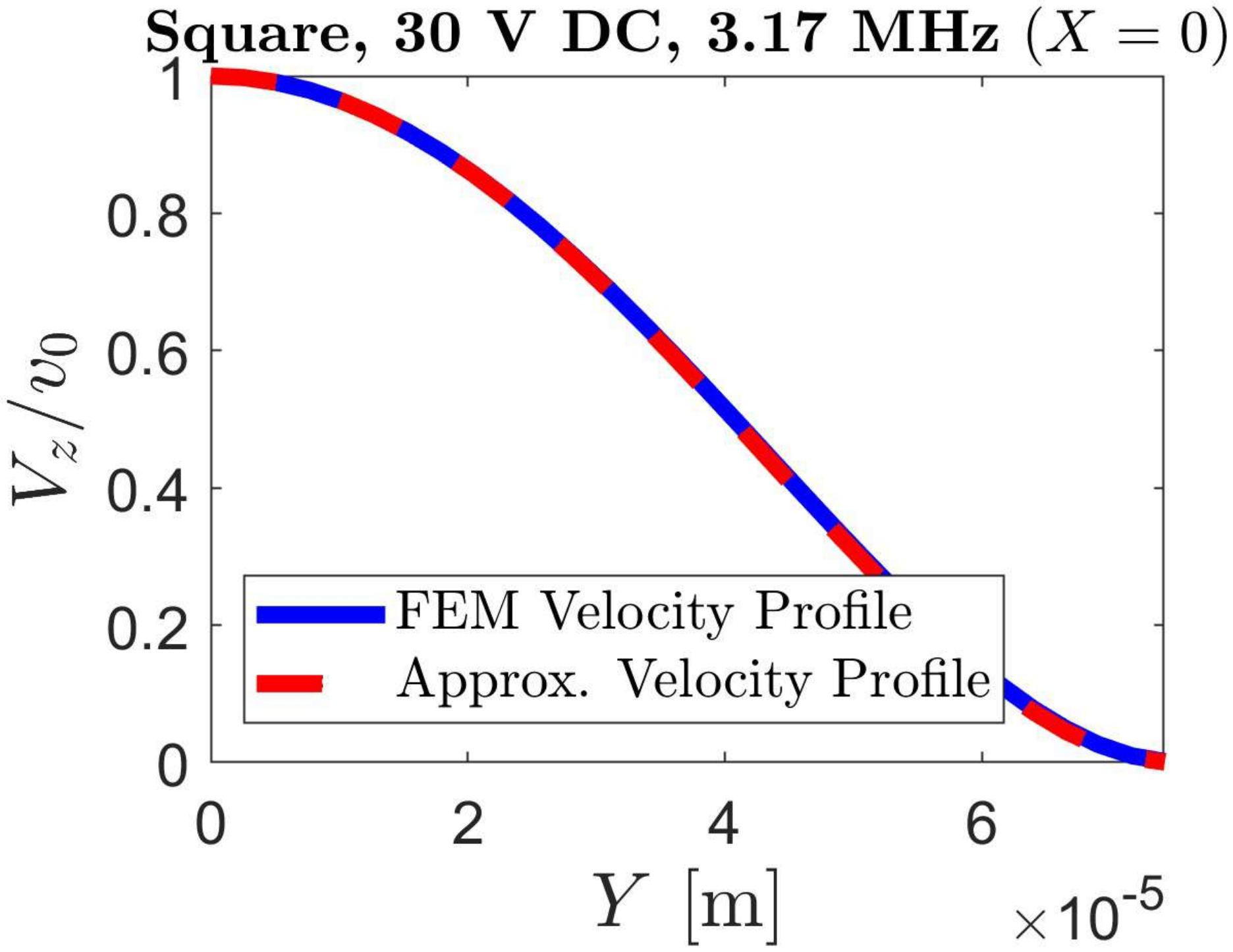}
\includegraphics[width=4.5cm]{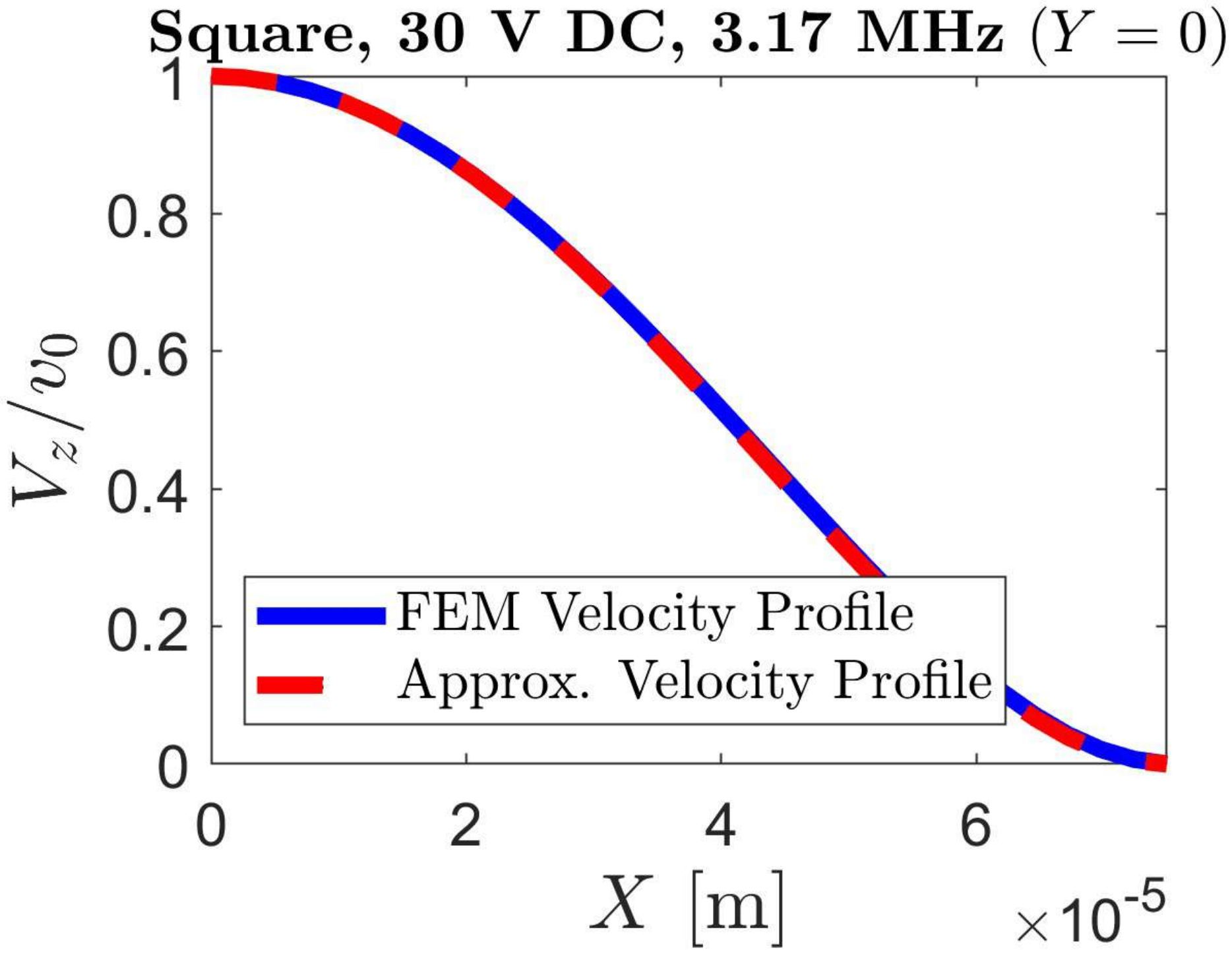}
\includegraphics[width=4.5cm]{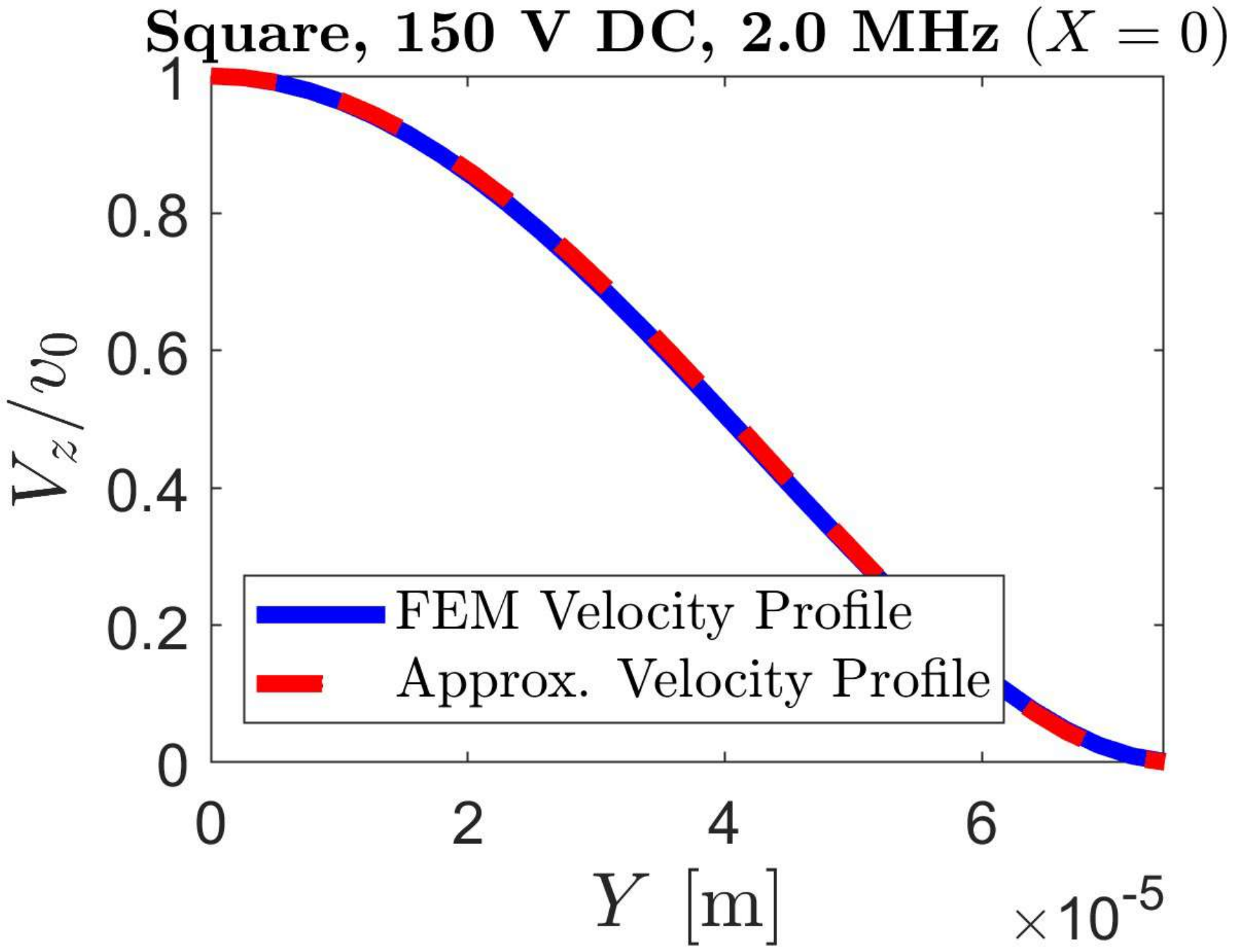}
\includegraphics[width=4.5cm]{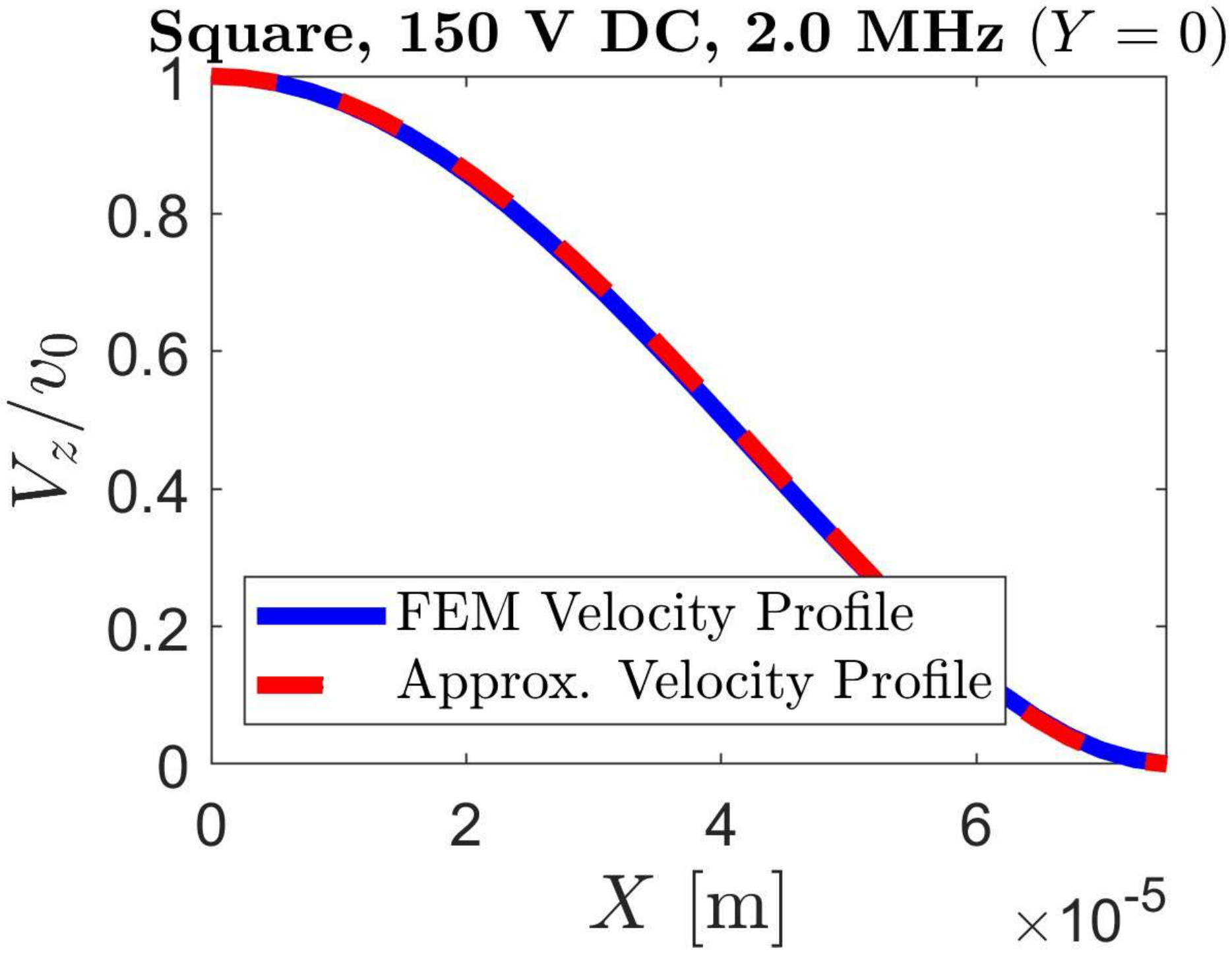}
\break
\includegraphics[width=4.5cm]{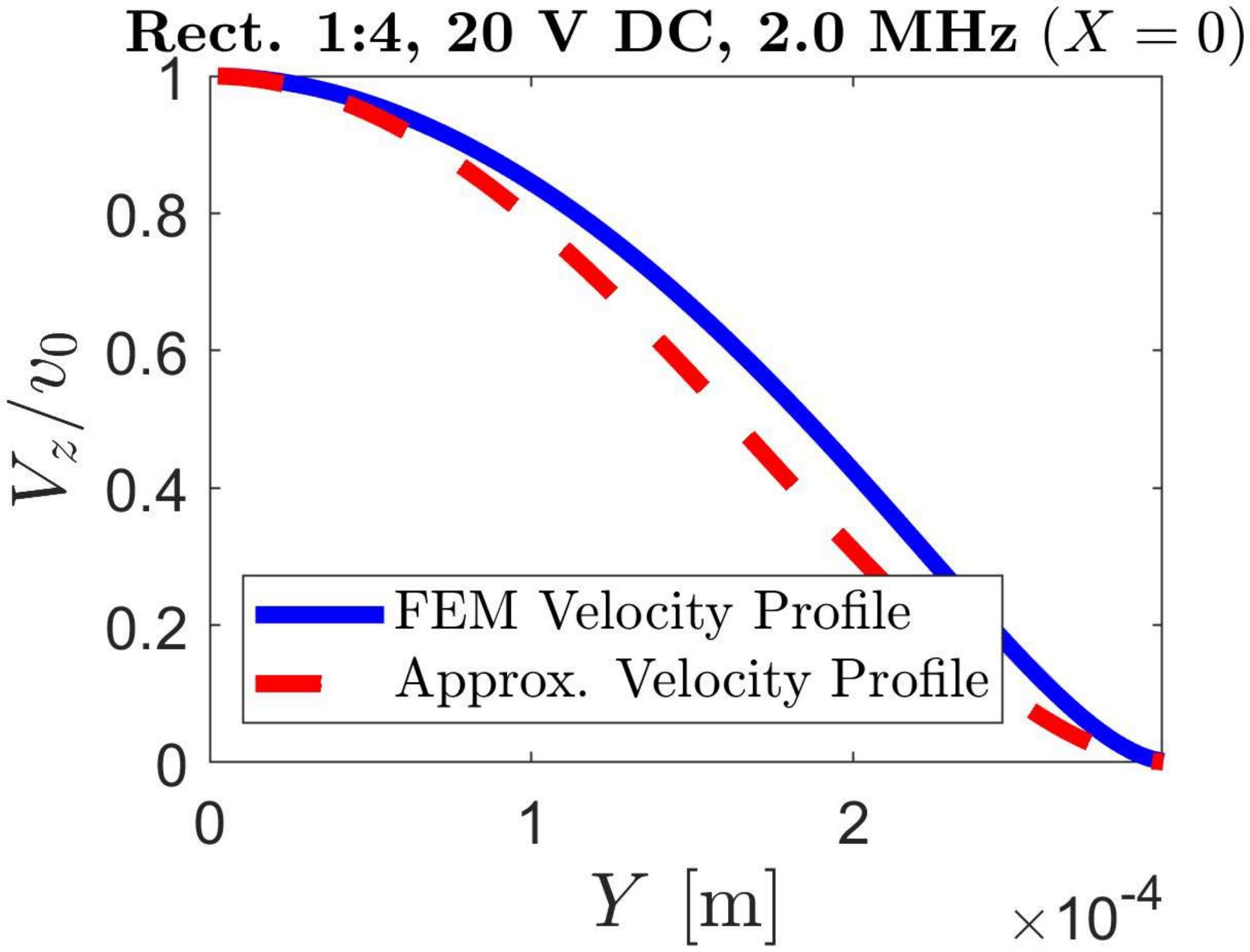}
\includegraphics[width=4.5cm]{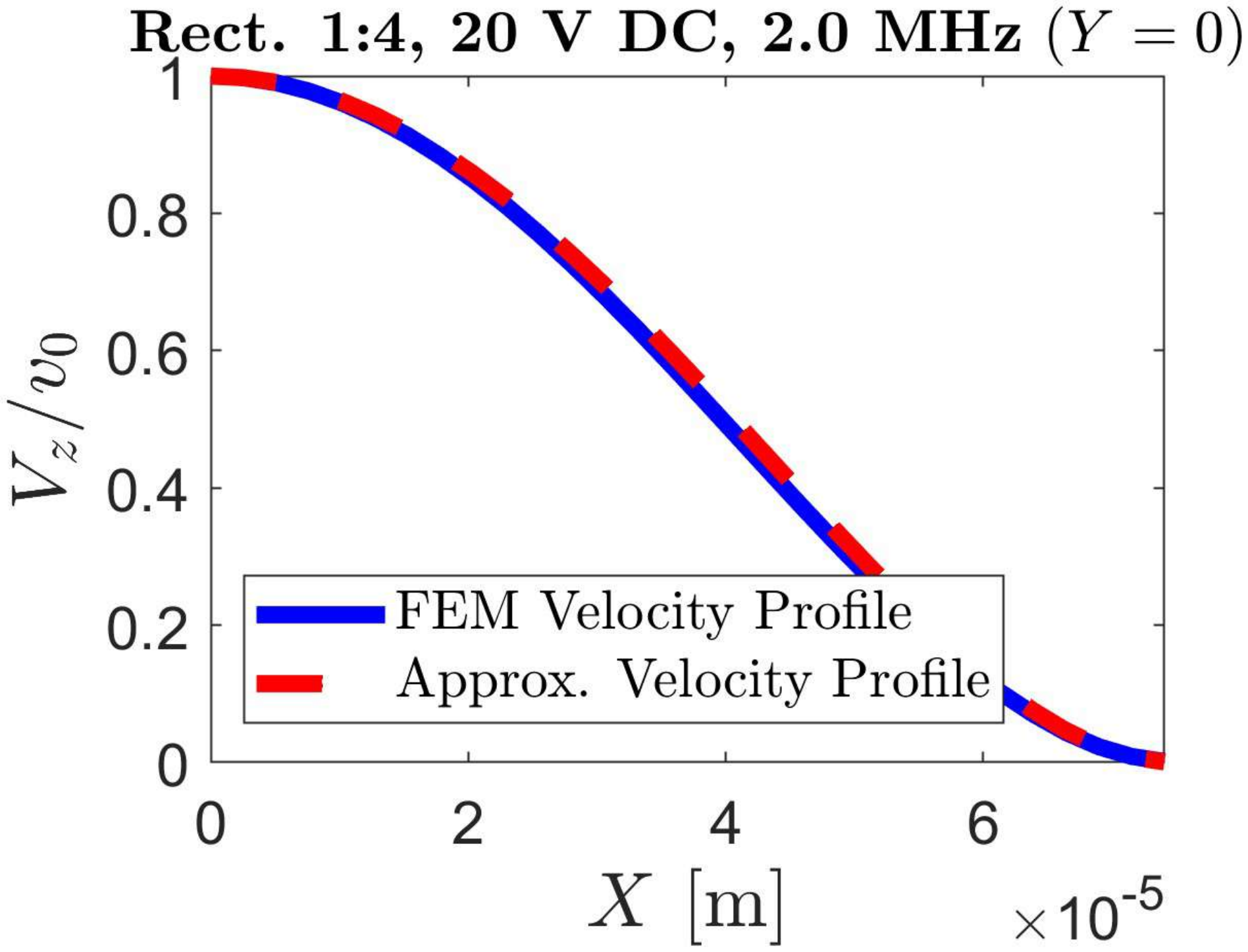}
\includegraphics[width=4.5cm]{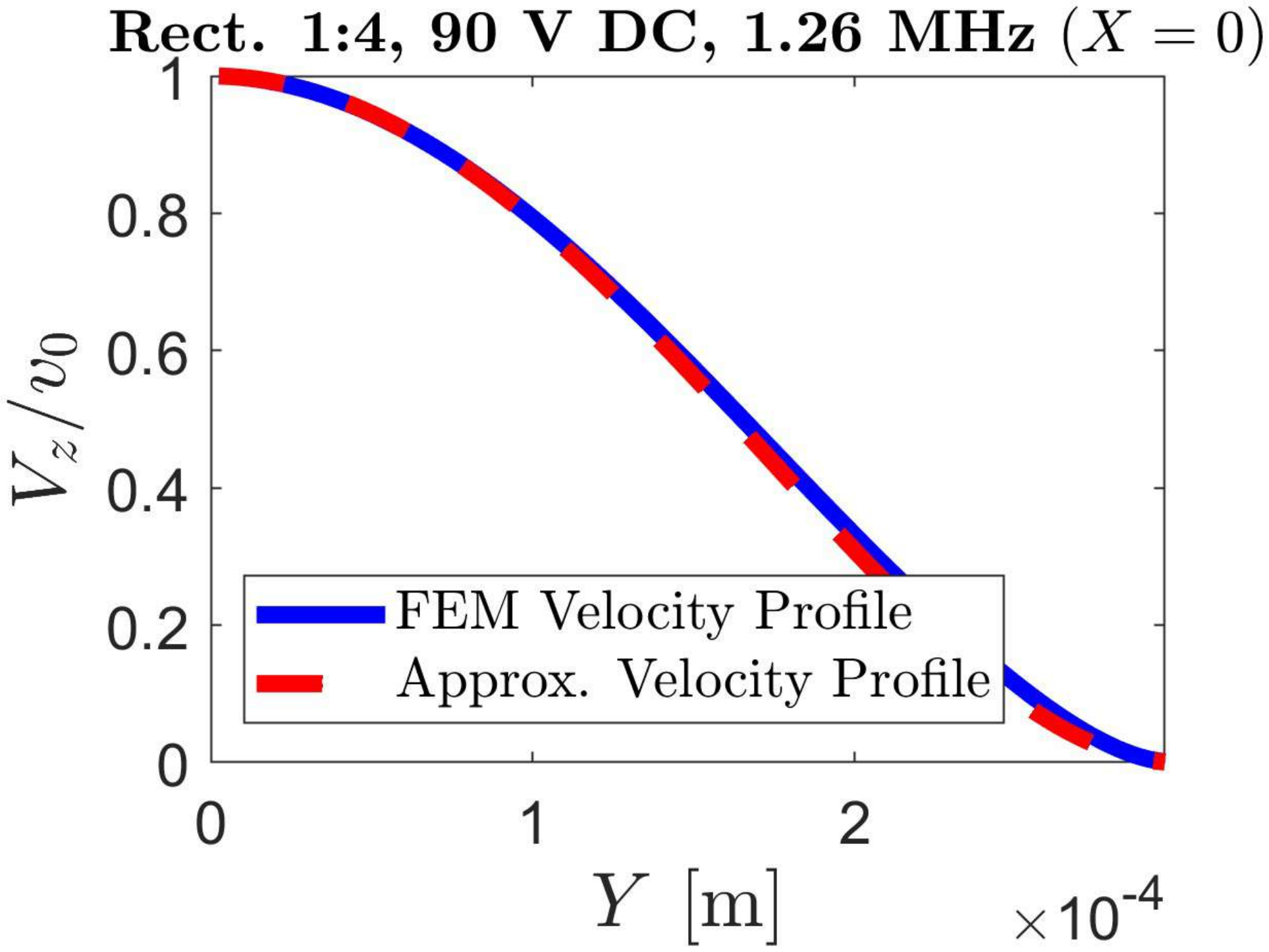}
\includegraphics[width=4.5cm]{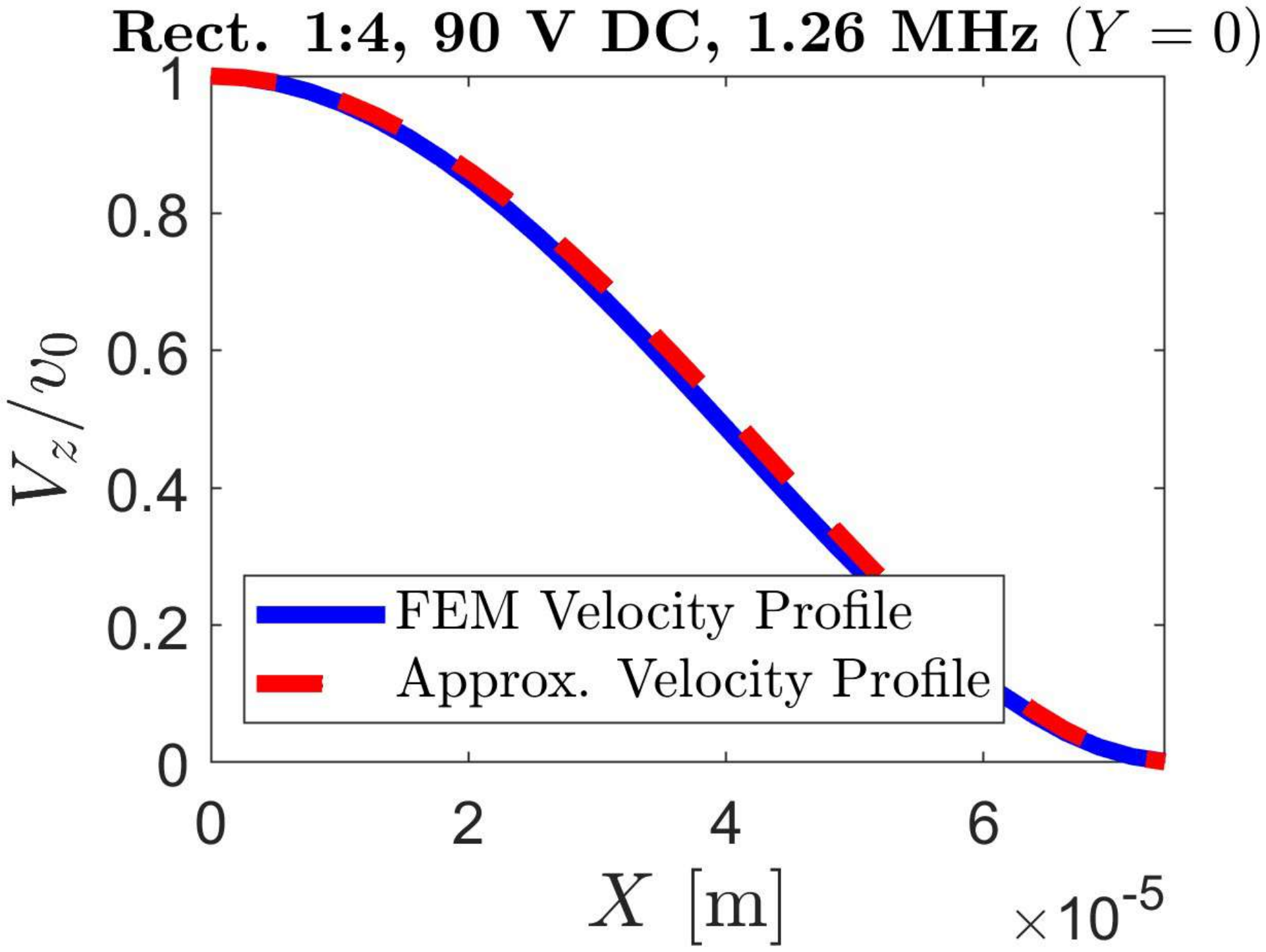}
\break
\includegraphics[width=4.5cm]{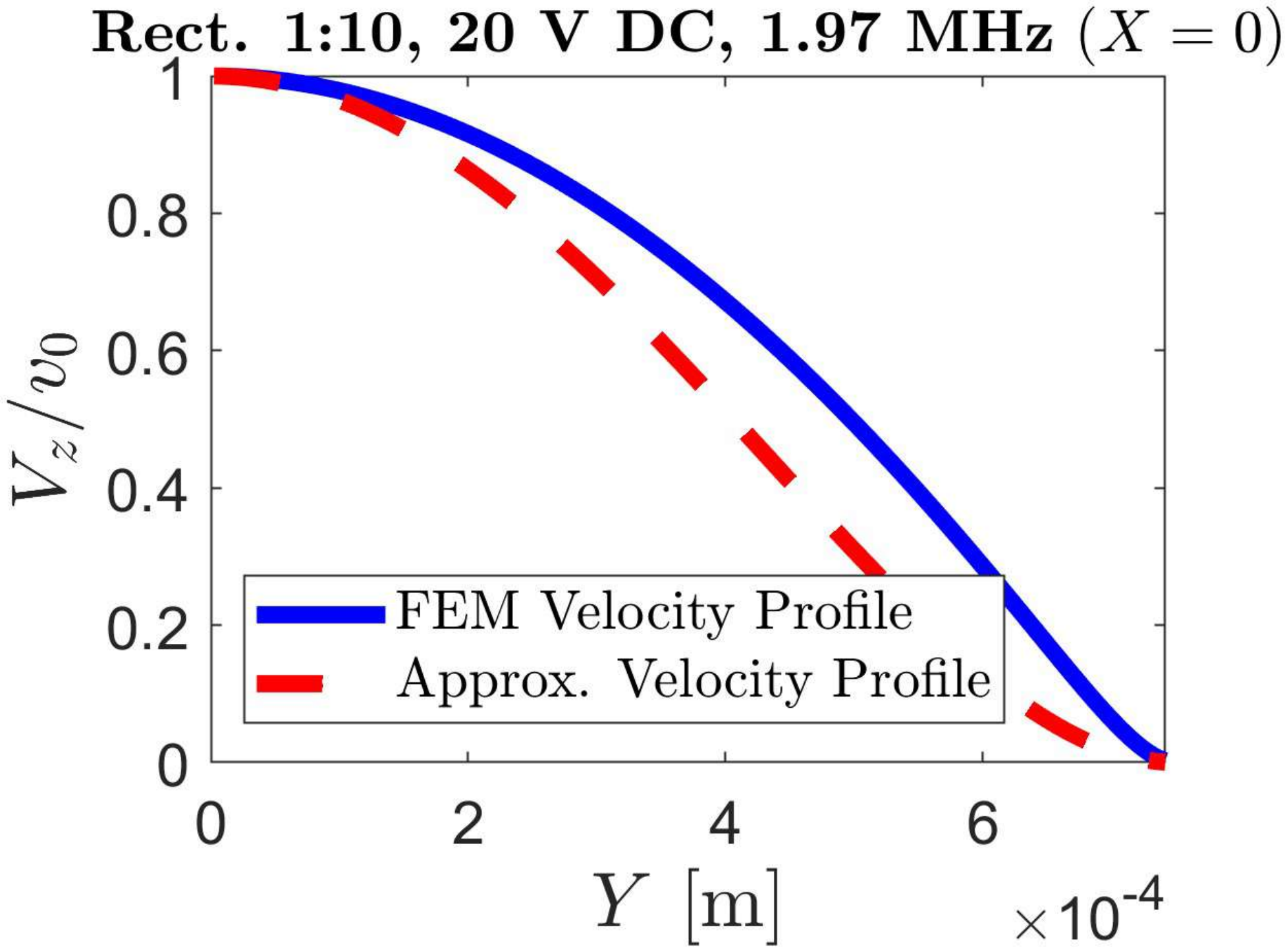}
\includegraphics[width=4.5cm]{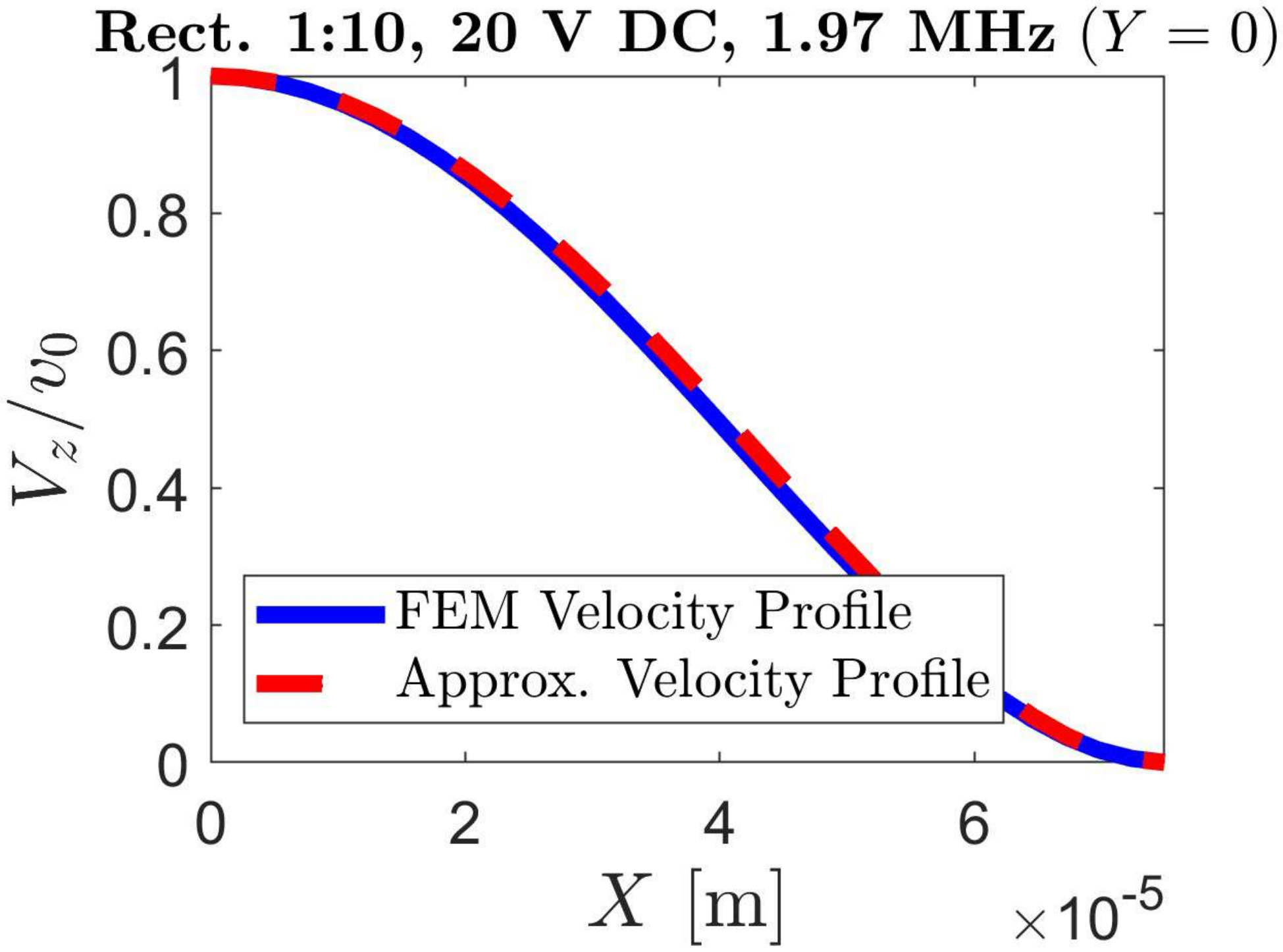}
\includegraphics[width=4.5cm]{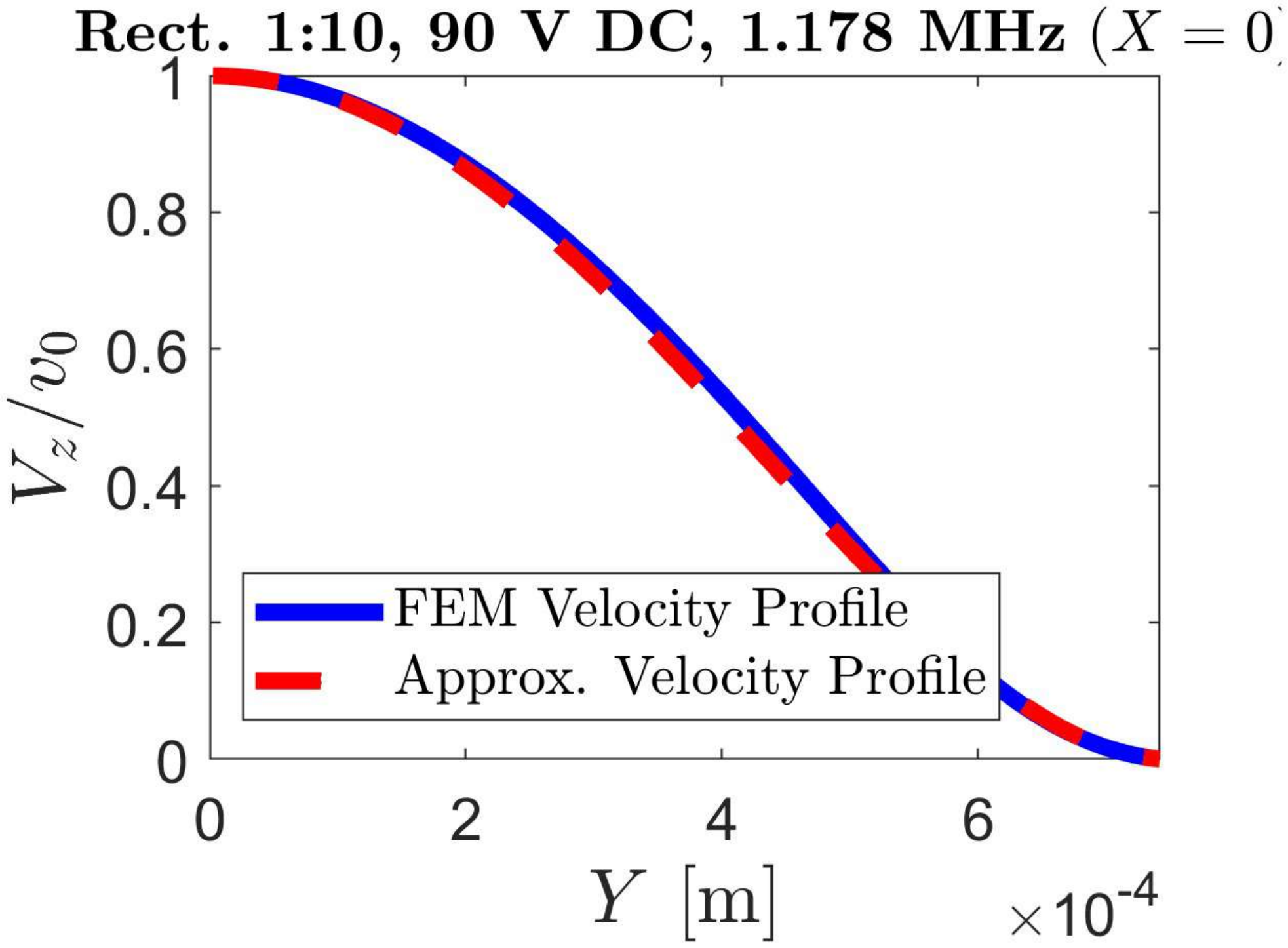}
\includegraphics[width=4.5cm]{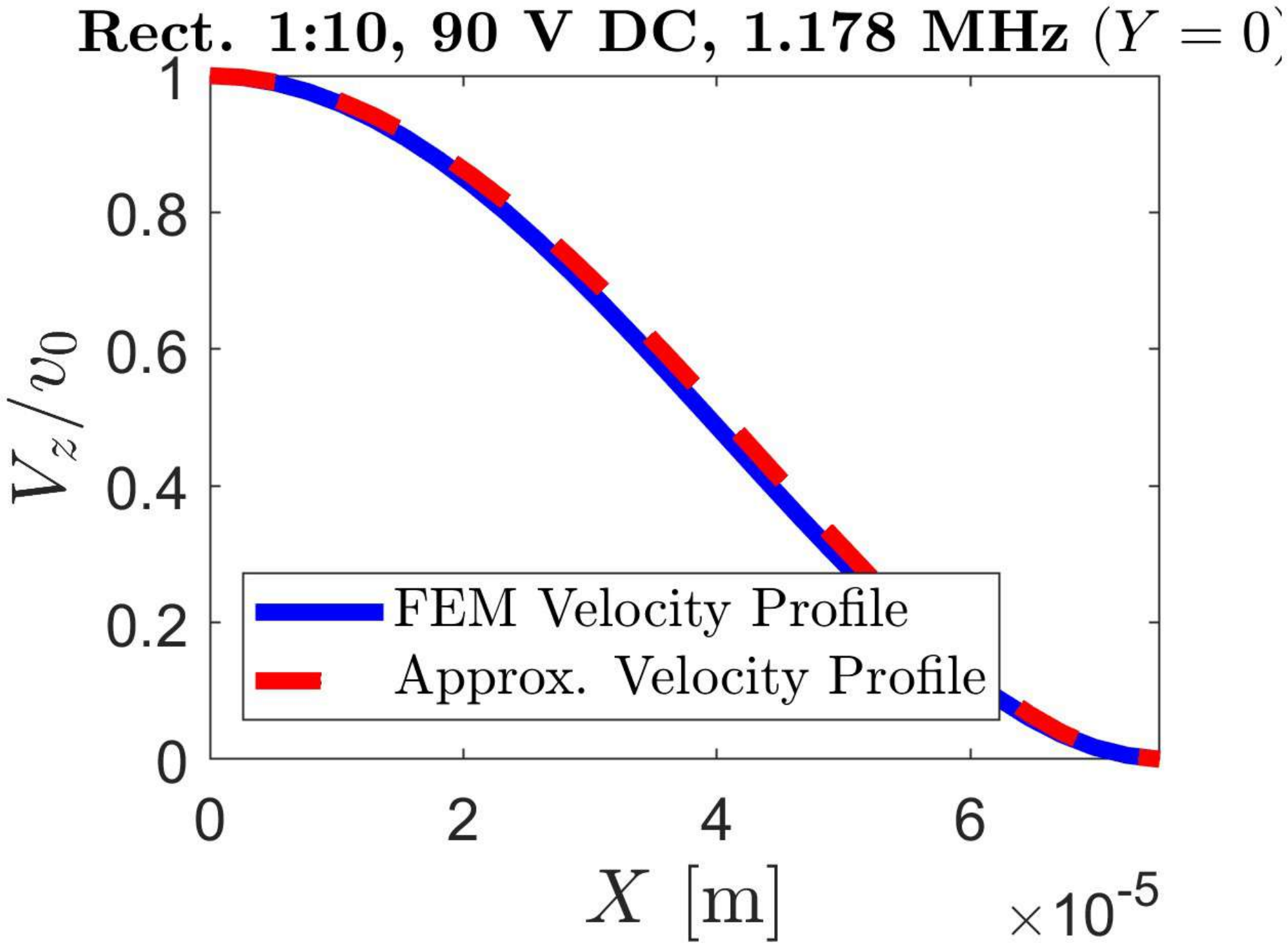}
\break
\includegraphics[width=4.5cm]{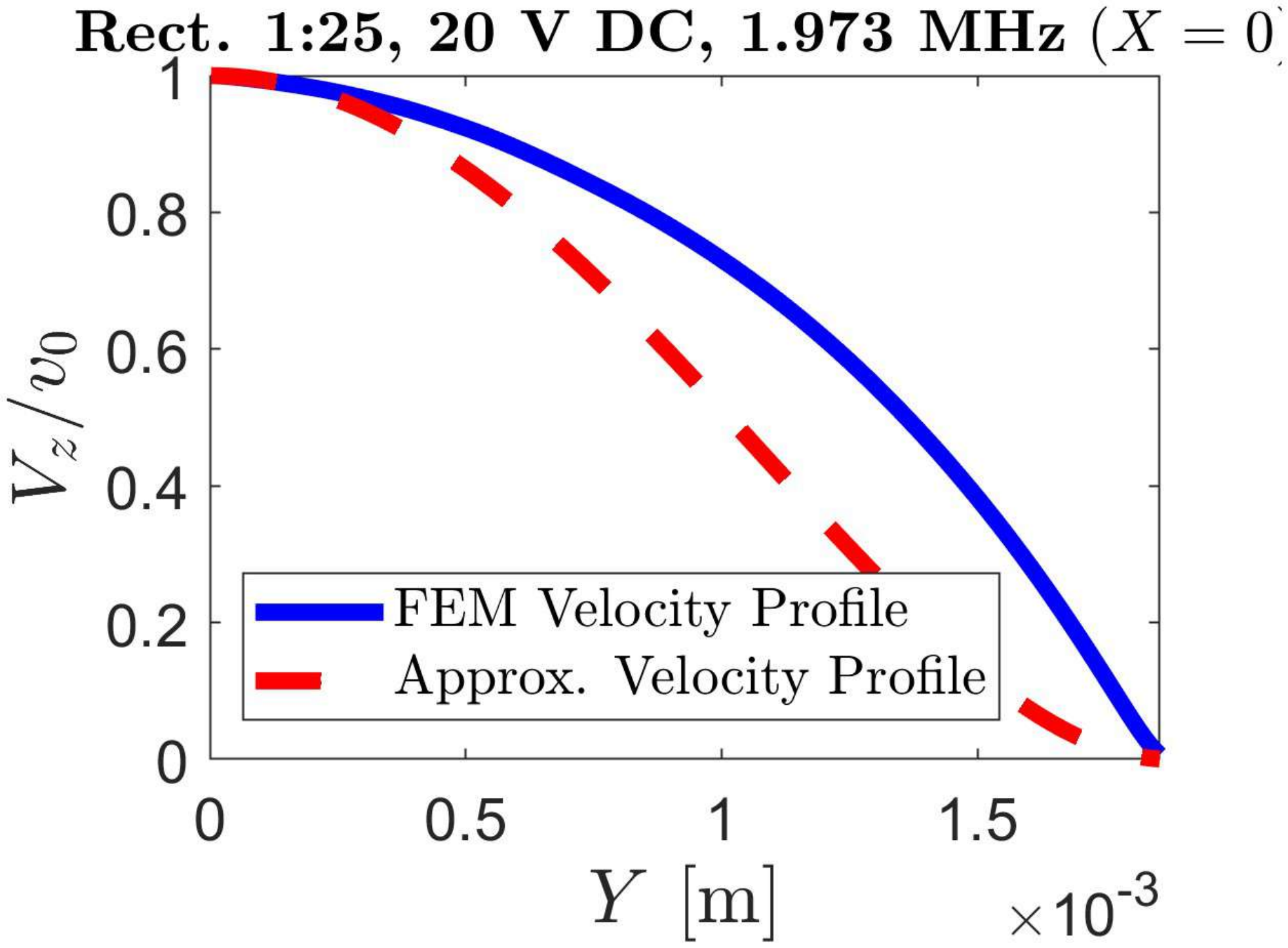}
\includegraphics[width=4.5cm]{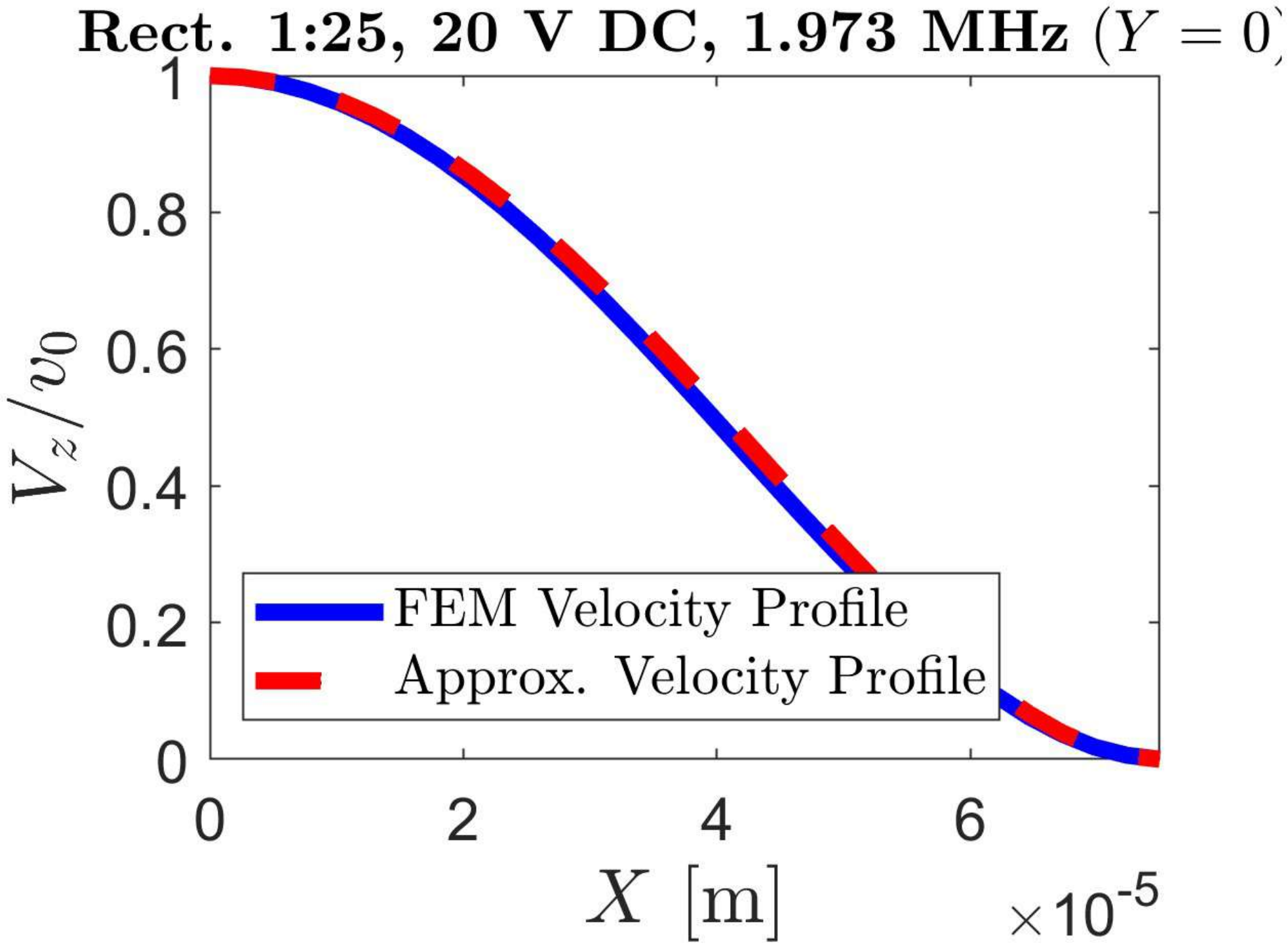}
\includegraphics[width=4.5cm]{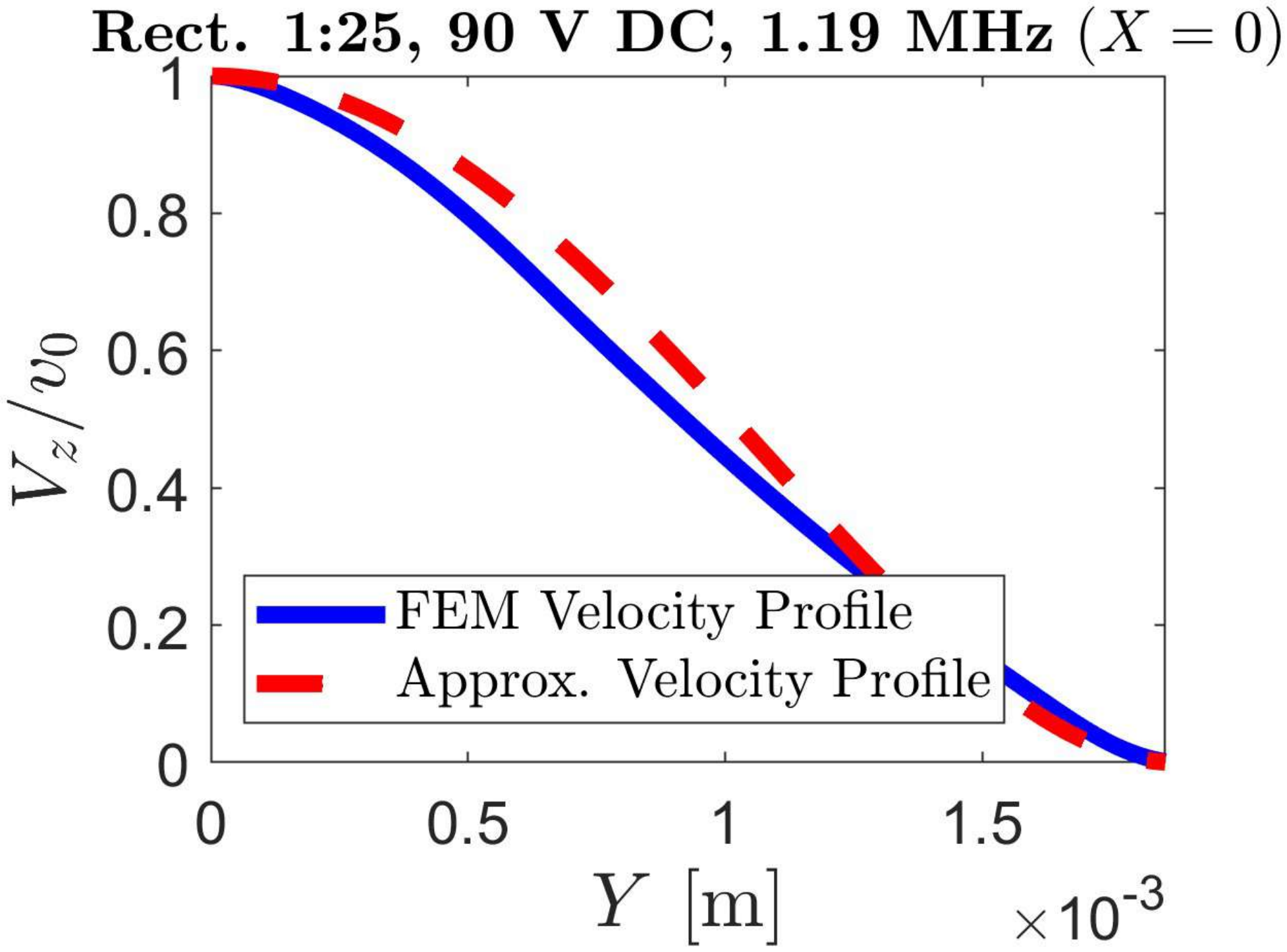}
\includegraphics[width=4.5cm]{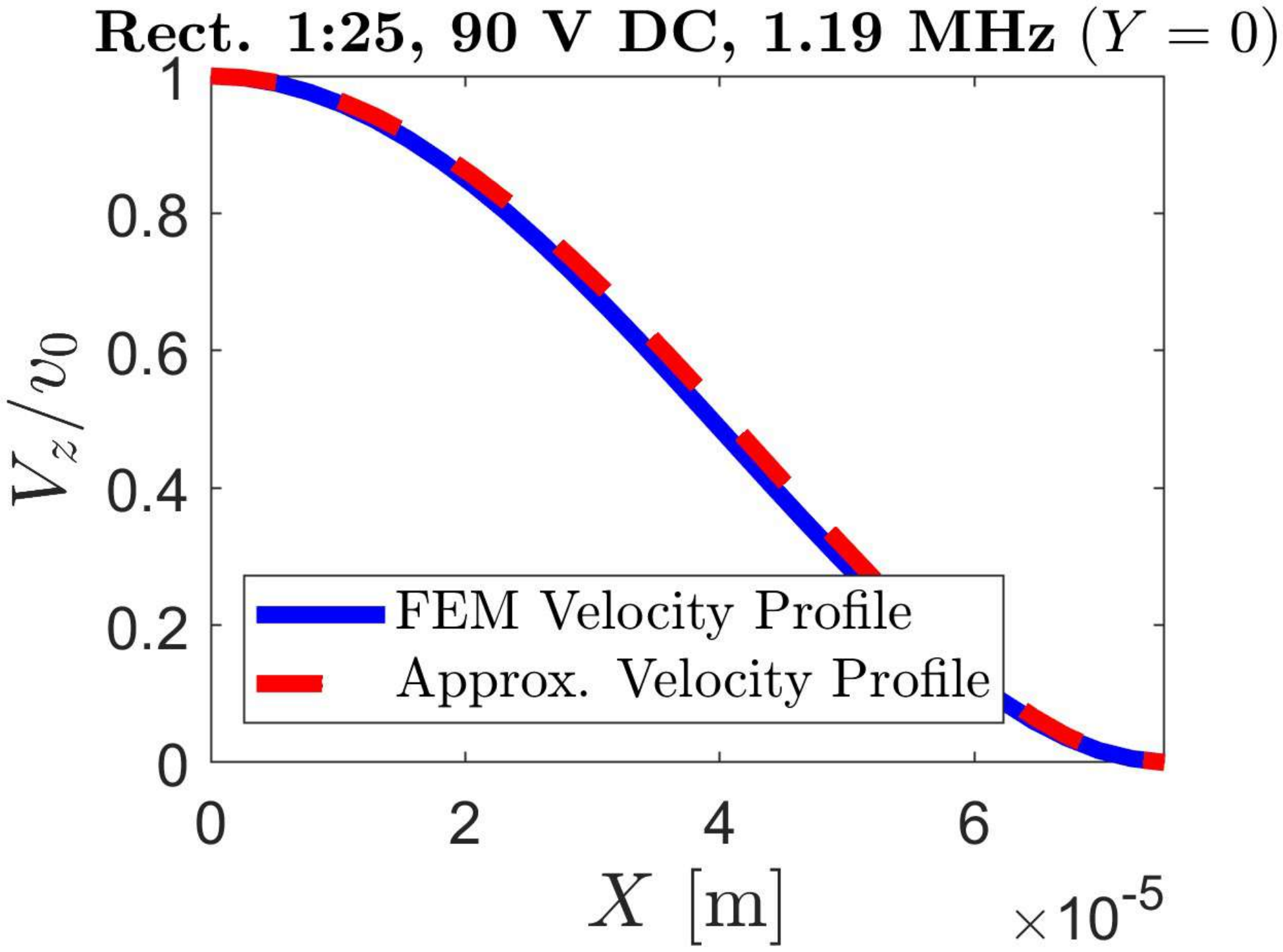}
\end{adjustwidth}
\caption{The cross-sectional velocity profiles from FEM and the proposed approximation of \eqref{v(x,y)} for the square, 1:4, 1:10, and 1:25 rectangular CMUTs. The FEM velocity profiles were extracted from frequencies near the CMUT's fundamental resonance at the given DC bias. For the rectangular CMUTs, their length was along the y-axis. Note the shape change of the cross-sectional velocity profile along the length as the aspect ratio grows from 4 to 25. The absolute relative errors of the approximate profile compared to the FEM are stated in Table~\ref{tab_ARE}. \label{all_vel_profs}}
\end{figure}

\begin{linenomath}
\begin{equation} \label{ARE}
    ARE = \frac{\iint_A |v_{FEM} - v(x,y)| dA}{\iint_A v_{FEM} dA},
\end{equation}    
\end{linenomath}
where $A$ represents the (image) area of the CMUT membrane on the x-y plane, which is $4ab$. The FEM velocity profile is given by $v_{FEM}$, while $v(x,y)$ is the approximate velocity profile from \eqref{v(x,y)}. In this calculation, both $v_{FEM}$ and $v(x,y)$ are normalized with respect to the peak velocity, and the discrete version of  \eqref{ARE} was used as the FEM velocity profile has a finite number of nodes. The $ARE$ values calculated for the aspect ratios of 1, 4, 10, and 25 are shown in Table~\ref{tab_ARE}. Each of the reported $ARE$ values corresponds to the fundamental mode of the membrane with the applied DC voltage. The simulated velocity profiles are compared with our polynomial approximation at the fundamental mode for each aspect ratio in Figure~\ref{all_vel_profs}.

\begin{table}[H] 
\caption{The absolute relative error ($ARE$) values achieved with the velocity profile of \eqref{v(x,y)} is reported in this table. The FEM velocity profiles were extracted at the observed resonance frequency corresponding to each DC bias. The bias was $\sim 20\%$ or $\sim 90\%$ of collapse voltage which is equivalent to \qty{30}{V} and \qty{150}{V} for the square CMUT and \qty{20}{V} and \qty{90}{V} for other geometries, respectively. \label{tab_ARE}}

\newcolumntype{C}{>{\centering\arraybackslash}X}
\begin{tabularx}{\textwidth}{C|CC|CC}
\toprule
\multicolumn{1}{c|}{} & \multicolumn{2}{c|}{Low DC bias} & \multicolumn{2}{c}{High DC bias} \\
Aspect ratio ($b/a$)	& $ARE$ & Frequency [\unit{MHz}]	& $ARE$ & Frequency [\unit{MHz}]\\
\midrule
1		& \qty{0.57}{\%} & \qty{3.17}{}			& \qty{1.39}{\%} & \qty{2.0}{}\\
4		& \qty{9.36}{\%} & \qty{2.0}{}			& \qty{3.05}{\%} & \qty{1.26}{}\\
10		& \qty{14.56}{\%} & \qty{1.97}{}			& \qty{2.33}{\%} & \qty{1.178}{}\\
25		& \qty{20.04}{\%} & \qty{1.973}{}			& \qty{10.45}{\%} & \qty{1.19}{}\\
\bottomrule
\end{tabularx}
\end{table}

The $ARE$ values in Table~\ref{tab_ARE}, and the cross-sections in Figure~\ref{all_vel_profs} both indicate that the our polynomial shape model more accurately models the velocity profile at low aspect ratios. As we show in Section \ref{derivation}, the radiation impedance depends on the spatial Fourier Transform of the velocity profile. Thus, one would expect that a more accurate approximation of the velocity profile would yield a more accurate approximation of the radiation impedance. Our polynomial shape model closely models the fundamental mode of lower aspect ratios; however, at high aspect ratios, agreement becomes worse along the long dimension.  This motivated us to investigate an alternative velocity model for high aspect ratios.

\subsubsection{Approximate Velocity Profile of Long Rectangular Membranes}

To address the potential lower accuracy of \eqref{v(x,y)} for long (high aspect ratio) rectangular membranes, we decided to explore another approximate velocity specifically used for long membranes. As shown in the left-most column plots of Figure~\ref{all_vel_profs}, as the aspect ratio increases, so does the average velocity along the long dimension. This hints that a mathematically simpler velocity profile may be worth considering for such membranes.

The 1D approximation assumes that the velocity profile over the aperture is only a function of x (coordinate along the width). This can be mathematically expressed by replacing the y-dependent half of \eqref{v(x,y)} with a rectangle function. This approximate velocity profile is given by \eqref{v1d} in the Appendix \ref{1dapprox}, along with the analytical derivations leading to the corresponding radiation impedance.

\section{Radiation Impedance Integral} \label{derivation}
Having explored the accuracy of our approximate velocity profile, it can now be used to compute the radiation impedance based on \eqref{v(x,y)}. From the Rayleigh-Sommerfeld equation for Neumann boundary conditions derived in \cite{cobbold}, we can write \eqref{press_1} for the pressure phasor. 

\begin{linenomath}
\begin{equation}\label{press_1}
    \tilde{p}(\textbf{r}) = j\rho c k \iint_{S_0} v(x_0,y_0) G(\textbf{R}) \,ds_0 ,
\end{equation}    
\end{linenomath}
where $G(\textbf{R})$ is the Green's function and $\textbf{R}=\textbf{r}-\textbf{r}_0$ is the vector from the point source to the point of observation. The expression for the free-space Green's function is shown in \eqref{green1} in its Euclidean form. 
\begin{linenomath}
\begin{equation}\label{green1}
    G(\textbf{R}) = \frac{e^{j\textbf{k}.\textbf{R}}}{2\pi R}
\end{equation}    
\end{linenomath}

In our calculation, we used an equivalent expression for the Green's function given as \eqref{green}. This form of the Green's function is commonly used \cite{TimMellow, fourierG, BERANEK2012}. 

\begin{linenomath}
\begin{equation}\label{green}
    G(\textbf{R}) = \frac{-j}{4\pi^2} \\ \int_{-\infty}^{\infty} \int_{-\infty}^{\infty} \frac{e^{-j(k_x(x-x_0) + k_y(y-y_0) + k_z(z-z_0))}}{k_z} \,dk_x\,dk_y ,
\end{equation}    
\end{linenomath}
where $k_z$ has a piece-wise definition stated below:
\begin{linenomath}
\begin{equation} \label{k_z}
    k_z = \Biggl\{ \begin{array}{cc} \sqrt{k^2 - k_x^2 - k_y^2} & k_x^2 + k_y^2 \le k^2 \\ -j\sqrt{k_x^2 + k_y^2 - k^2} & k_x^2 + k_y^2 > k^2 \\ \end{array} 
\end{equation}    
\end{linenomath}

Using \eqref{green} in \eqref{press_1}, and replacing $v(x_0,y_0)$ with \eqref{v(x,y)}, we can move the inner integral to the outside. Then, the pressure phasor can be expressed as \eqref{press_2} where $z_0=0$ since the velocity profile describes the particle velocity on the x-y plane.
\begin{adjustwidth}{-\extralength}{0cm}
\begin{equation}\label{press_2}
    \tilde{p}(\textbf{r}) = \frac{\rho c k v_0}{4\pi^2} \iint_{-\infty}^{\infty} \frac{e^{-j(k_xx + k_yy + k_zz)}}{k_z} \\ \left[ \iint_{S_0} \left( 1 - \left(\frac{x_0}{a}\right)^2 \right)^2 \left( 1 - \left(\frac{y_0}{b}\right)^2 \right)^2 e^{j(k_xx_0+k_yy_0)} \,dx_0\,dy_0 \right] \,dk_x\,dk_y
\end{equation}    
\end{adjustwidth}

Note that the inner integral in \eqref{press_2} is a spatial Fourier transform of the velocity profile. For the sake of brevity, we abbreviate the result of the inner integral of \eqref{press_2} as $a\ S(k_xa)\times b\ S(k_yb)$, with $S(k_xa)$ defined as \eqref{Sx}.

\begin{linenomath}
\begin{equation}\label{Sx}
     S(k_xa) = \frac1a \int_{-a}^{a} e^{ik_xx_0} \left[1-(\frac{x_0}{a})^2\right]^2 \,dx_0 = -\frac{16 \left( 3 k_xa \cos (k_xa) + \left(k_x^2 a^2 -3\right) \sin (k_xa)\right)}{k_x^5 a^5} ,
\end{equation}    
\end{linenomath}

With an expression for the pressure phasor obtained, the radiation impedance can now be calculated by evaluating \eqref{powerdef}.
\begin{linenomath}
\begin{equation} \label{powerdef}
    P_{tot} = \iint_S \tilde{p}(\textbf{r}) v^*_n(\textbf{r}) \,ds ,
\end{equation}
\end{linenomath}
where the surface of integration is chosen to be the x-y plane again. Using the expression for pressure from \eqref{press_2} while setting $z=0$ yields the familiar integral of  \eqref{power1}. As all points on the membrane are assumed to be in phase with each other (in the fundamental mode of oscillation), the velocity phasor is real, and thus the conjugate is identical.

\begin{linenomath}
\begin{equation}\label{power1}
    P_{tot} = \iint_{S_0} v_0 \left( 1 - \left(\frac{x}{a}\right)^2 \right)^2 \left( 1 - \left(\frac{y}{b}\right)^2 \right)^2 \ \frac{\rho c k a b v_0}{4\pi^2} \\
    \left[ \iint_{-\infty}^{\infty} \frac{e^{-j(k_xx + k_yy)}}{k_z} S(k_xa)\ S(k_yb) \,dk_x\,dk_y \right] \,ds ,
\end{equation}    
\end{linenomath}
changing the order of integration, we obtain \eqref{power2} utilizing \eqref{Sx}.
\begin{linenomath}
    \begin{equation}\label{power2}
    P_{tot} = \frac{\rho c k a^2 b^2 v_0^2}{4\pi^2} \iint_{-\infty}^{\infty} S^2(k_xa)\ S^2(k_yb) \frac{\,dk_x\,dk_y}{k_z} 
\end{equation}
\end{linenomath}

This relationship between the total power emitted and the square of the spatial Fourier Transform of the velocity profile was originally shown by C. J. Bouwkamp \cite{bouwkamp}, through the introduction of the far-field directivity function. To simplify this integral for total power, we perform a standard change of variables used in rectangular geometries \cite{TimMellow}. Further details about this procedure are provided in Appendix \ref{changeVars}. Separating the real and complex portions, we obtain expressions for $R_R$ and $X_R$, the real and imaginary parts, respectively. Note that the bounds in \eqref{impR} and \eqref{impI} are more tractable than in \eqref{power2} due to the change of variables. All the expressions in this work were derived using Wolfram Mathematica.

\begin{adjustwidth}{-\extralength}{0cm}
    \begin{equation}\label{impR}
    R_R = \frac{\rho c k^2 a^2 b^2}{4 \pi^2} \ \frac{99225}{16384} \\ \int_{0}^{2\pi} \int_{0}^{1} S^2(kta\cos\phi)\ S^2(ktb\sin\phi) \frac{t\,dt\,d\phi}{\sqrt{1-t^2}}
\end{equation}
\end{adjustwidth}

\begin{adjustwidth}{-\extralength}{0cm}
    \begin{equation}\label{impI}
    X_R = - \frac{\rho c k^2 a^2 b^2}{4 \pi^2} \ \frac{99225}{16384} \\ \int_{0}^{2\pi} \int_{1}^{\infty} S^2(kta\cos\phi)\ S^2(ktb\sin\phi) \frac{t\,dt\,d\phi}{\sqrt{t^2-1}}
    \end{equation}
\end{adjustwidth}

These expressions for radiation impedance were calculated using the RMS velocity as the reference velocity. The definition for RMS velocity is given in \eqref{VRMS}, alongside its evaluation using \eqref{v(x,y)}.

\begin{linenomath}
\begin{equation}\label{VRMS}
    V_\mathrm{RMS}^2 = \frac{\iint_A |v(\textbf{r})|^2 dA}{\iint_A dA} = \frac{16384}{99225} v_0^2
\end{equation}    
\end{linenomath}

Although initial attempts at numerically integrating \eqref{impR} and \eqref{impI} with Wolfram Mathematica did not converge, this was resolved by further investigating the behavior of the integrand. The expression \eqref{Sx} appears to diverge as its argument approaches zero, however, the convergence of this function can be verified by calculating the full Laurent series, and verifying that there are no terms that diverge at zero. The apparent divergence near zero is likely caused by numerical precision errors. Thus, accurate numerical computation of the integrals in \eqref{impR} and \eqref{impI} can be achieved by using Taylor expansions of $S(k_xa)$ and $S(k_yb)$ near zero in place of their original form.

Our MATLAB script for numerically integrating the radiation impedance is available at \cite{github_repo}. The direct output of the script is dimensionless, meaning the expressions of \eqref{impR} and \eqref{impI} were first divided by $4 ab \rho c$ before numerical integration. 

%%%%%%%%%%%%%%%%%%%%%%%%%%%%%%%%%%%%%%%%%%
\section{Results}
\subsection{Circular CMUT}
To contextualize our results with rectangular membranes, we began by comparing FEM simulations of circular CMUTs to the well-known analytical expressions for clamped radiators derived in \cite{martin_greenspan}. The circular CMUT was simulated using the same approach as the mentioned rectangular CMUTs, with a membrane thickness of \qty{4.9}{\micro\m} and a diameter of \qty{150}{\micro\m}. These dimensions were selected to be consistent with the simulated square CMUT.

\begin{figure}[H]
\begin{adjustwidth}{-\extralength}{0cm}
\centering
\includegraphics[width=8cm]{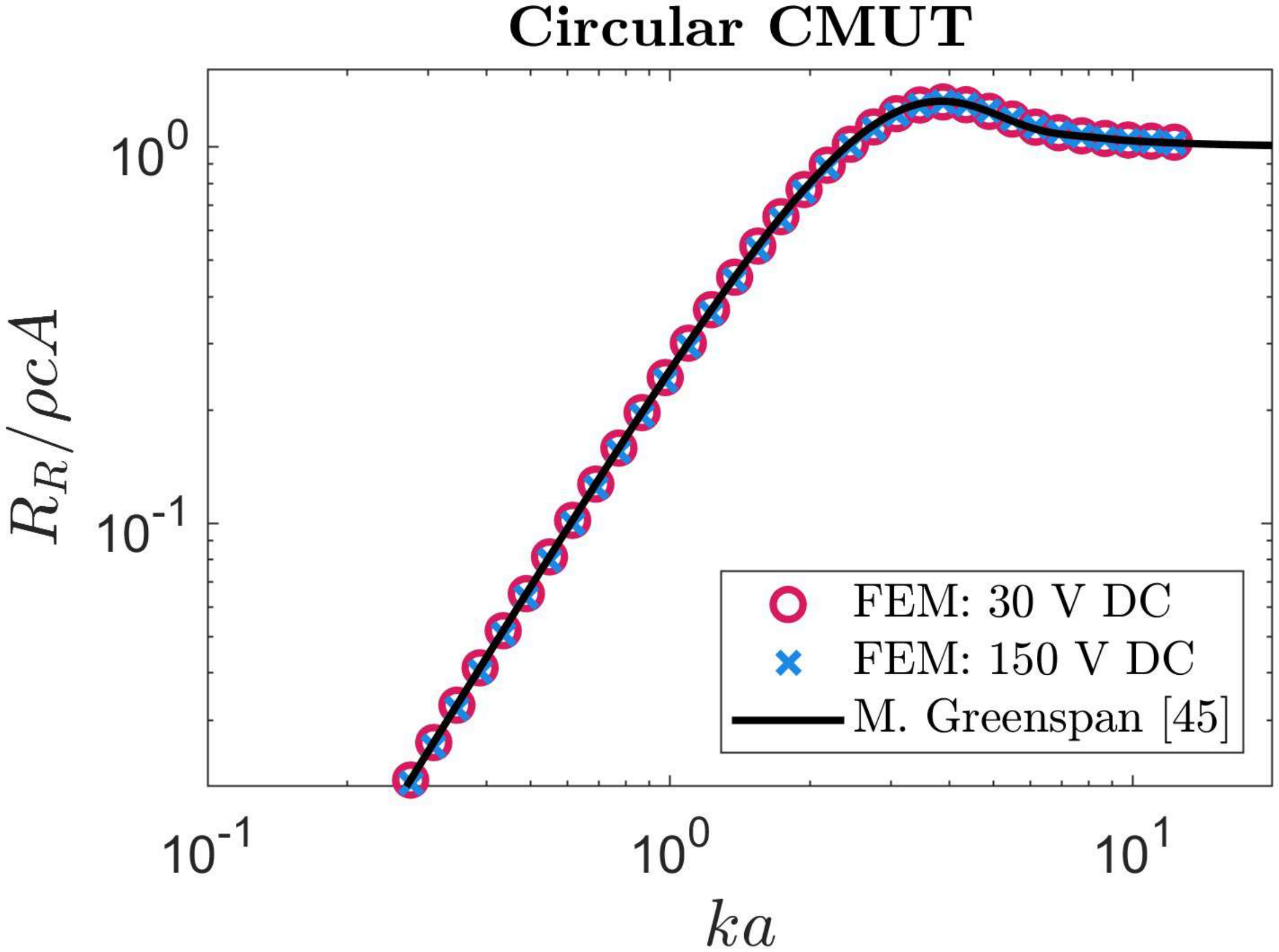}
\includegraphics[width=8cm]{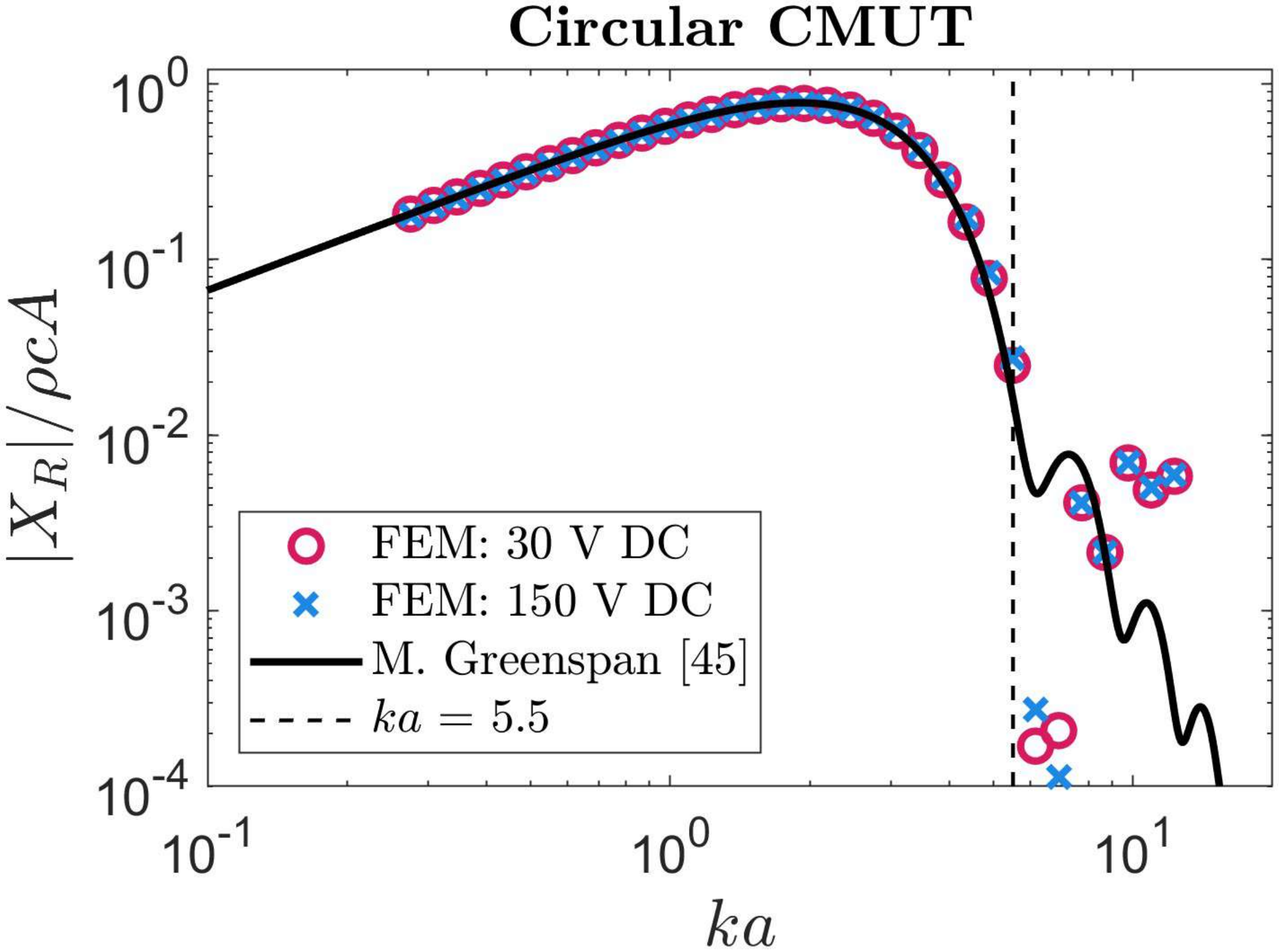}
\end{adjustwidth}
\caption{The radiation impedance of a circular CMUT obtained from simulation compared to the expression derived by M. Greenspan \cite{martin_greenspan}. The error of the theoretical predictions compared to FEM are quantified in Table~\ref{tab_MAPE}. \label{circ_ZR}}
\end{figure}

These circular membrane expressions are also a close analogue to our results, as they were derived assuming an approximate velocity profile. Similar to our approach, this approximate velocity profile was a fourth-order polynomial expression representing the fundamental mode of vibration. These analytical radiation impedance results are compared to our FEM simulations in  Figure~\ref{circ_ZR}. In the case of circular radiators, $a$ is taken to be the radius of the membrane, in contrast to our usual definition. 

As illustrated in Figure~\ref{circ_ZR}, the real part of the impedance closely matches between the analytical approach and FEM simulations at both high and low bias voltages. However, at high frequencies, the reactance began to diverge. Subsequent simulations were performed with finer meshes and varied acoustic media to investigate this divergence. Based on those simulations, our interpretation is that the primary sources of this divergence are higher-order modes not considered by the analytical method and medium attenuation. 

Based on these considerations, we defined a threshold of $ka=5.5$, below which the analytical reactance predictions closely matched FEM. It should be noted that for the purpose of optimizing CMUT designs, this limitation is not an important impediment as most designs occur below $ka=5.5$ \mbox{\cite{koymen, improved_koymen}}. Similarly, in some studies, the results above analogous $ka$ values are not reported for radiation resistance and impedance \mbox{\cite{Dingguo2009}}. Regardless, in this region, the mean absolute error (MAE) of the reactance was $0.0037 \rho cA$ when compared with FEM simulations at a low bias voltage and $0.0074 \rho cA$ at a high bias voltage. Whereas, across all frequencies, the $MAE$ of the radiation resistance was $0.0033 \rho cA$ and $0.0069 \rho cA$ when compared with FEM simulations at low and high bias voltages, respectively.

\subsection{Rectangular CMUT}
\begin{figure}[H]
\begin{adjustwidth}{-\extralength}{0cm}
\centering
\includegraphics[width=8cm]{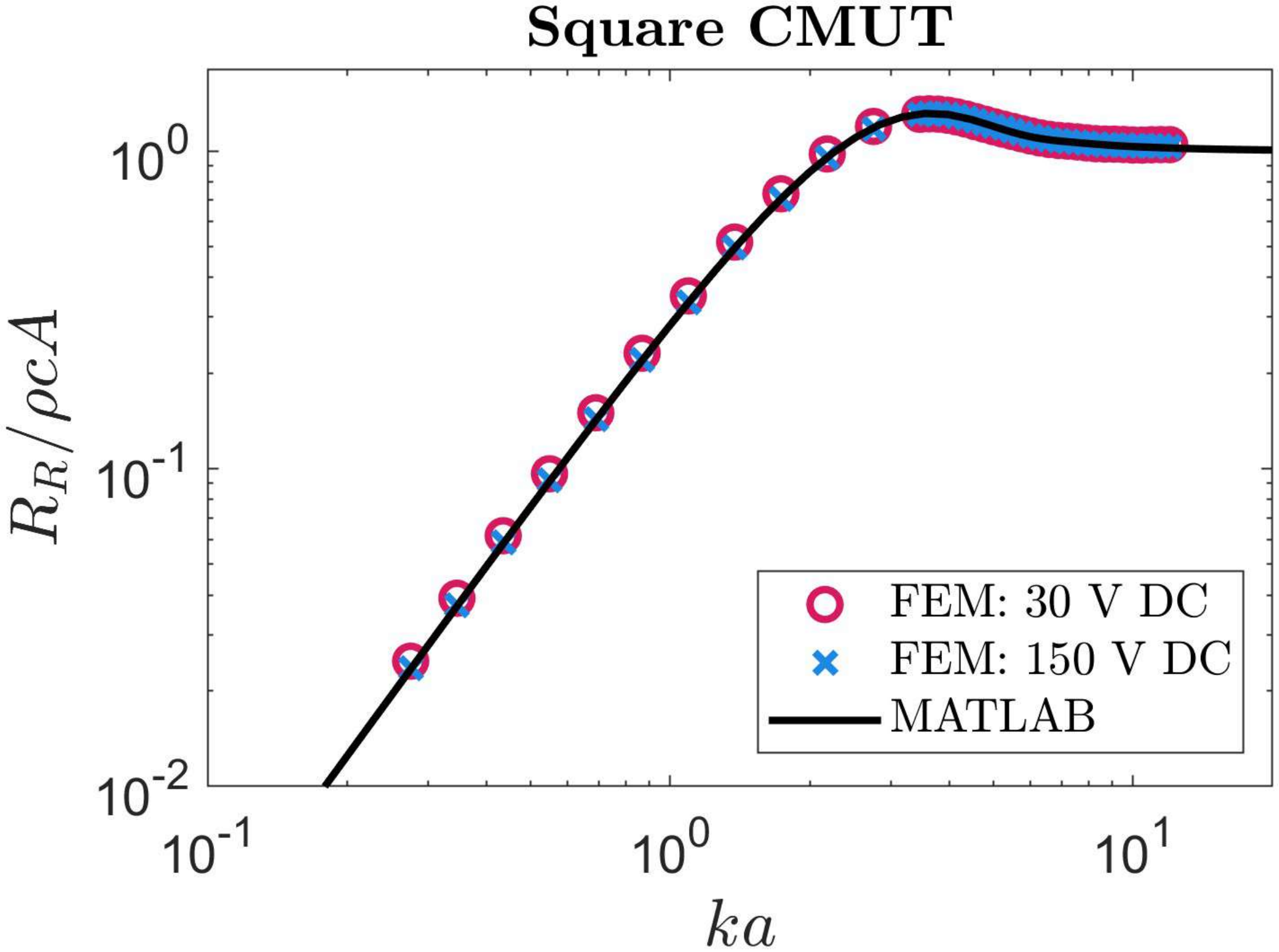}
\includegraphics[width=8cm]{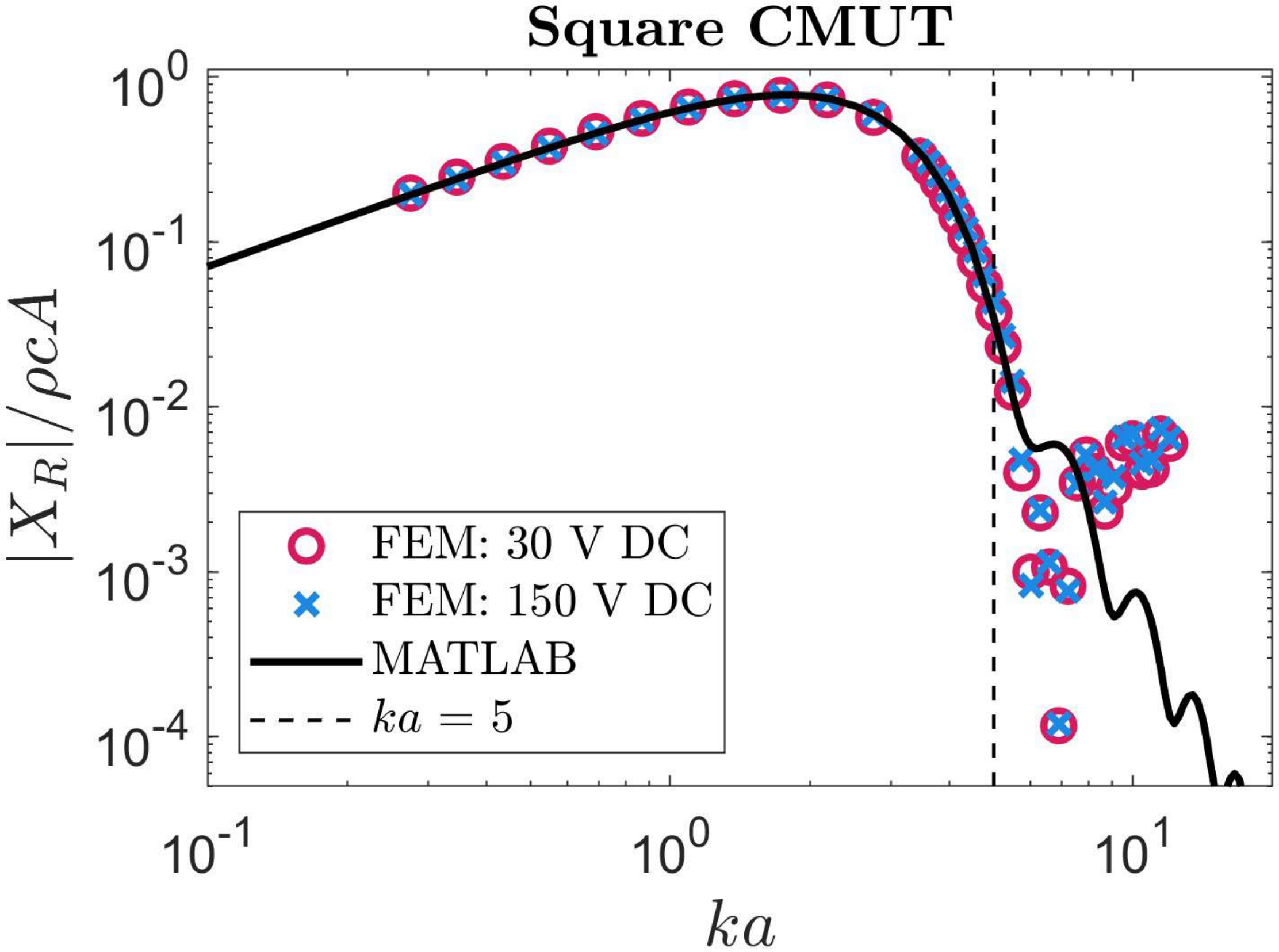}
\end{adjustwidth}
\caption{The radiation impedance of a square CMUT extracted from FEM compared to our method. \label{sq_ZR}}
\end{figure}

\begin{figure}[H]
\begin{adjustwidth}{-\extralength}{0cm}
\centering
\includegraphics[width=8cm]{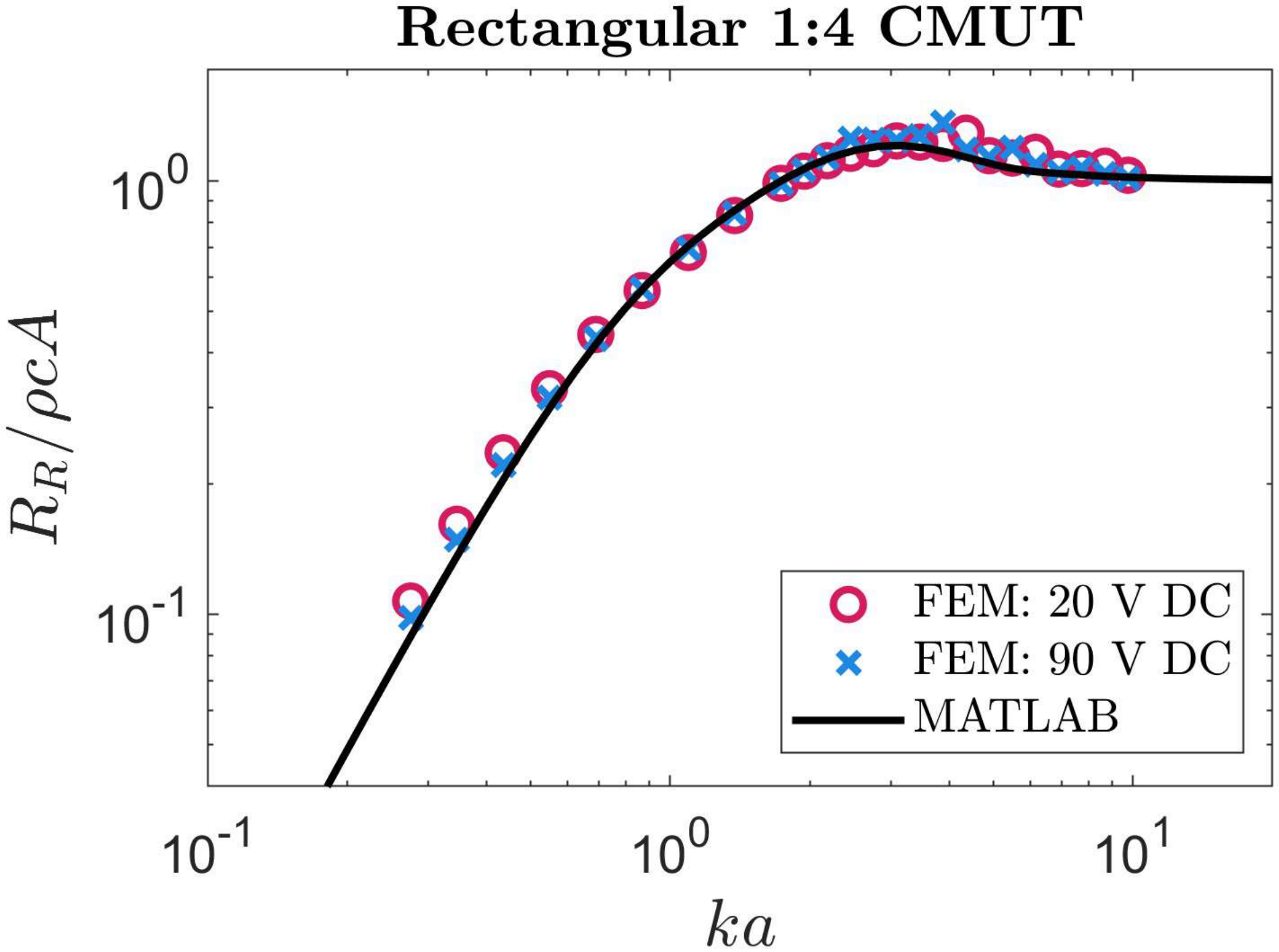}
\includegraphics[width=8cm]{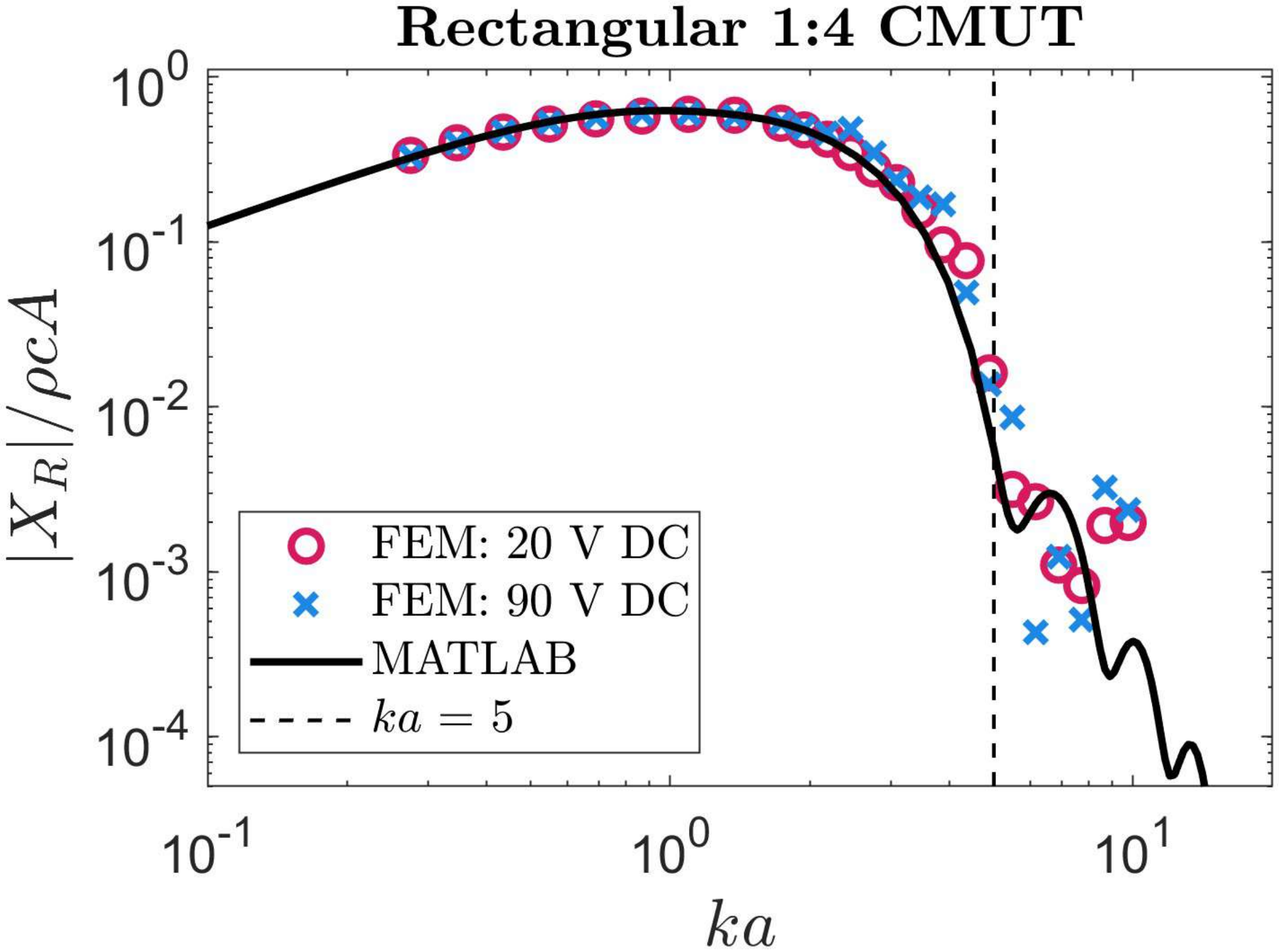}
\end{adjustwidth}
\caption{The radiation impedance of a rectangular 1:4 CMUT extracted from FEM compared to our method. \label{r4_ZR}}
\end{figure}

\begin{figure}[H]
\begin{adjustwidth}{-\extralength}{0cm}
\centering
\includegraphics[width=8cm]{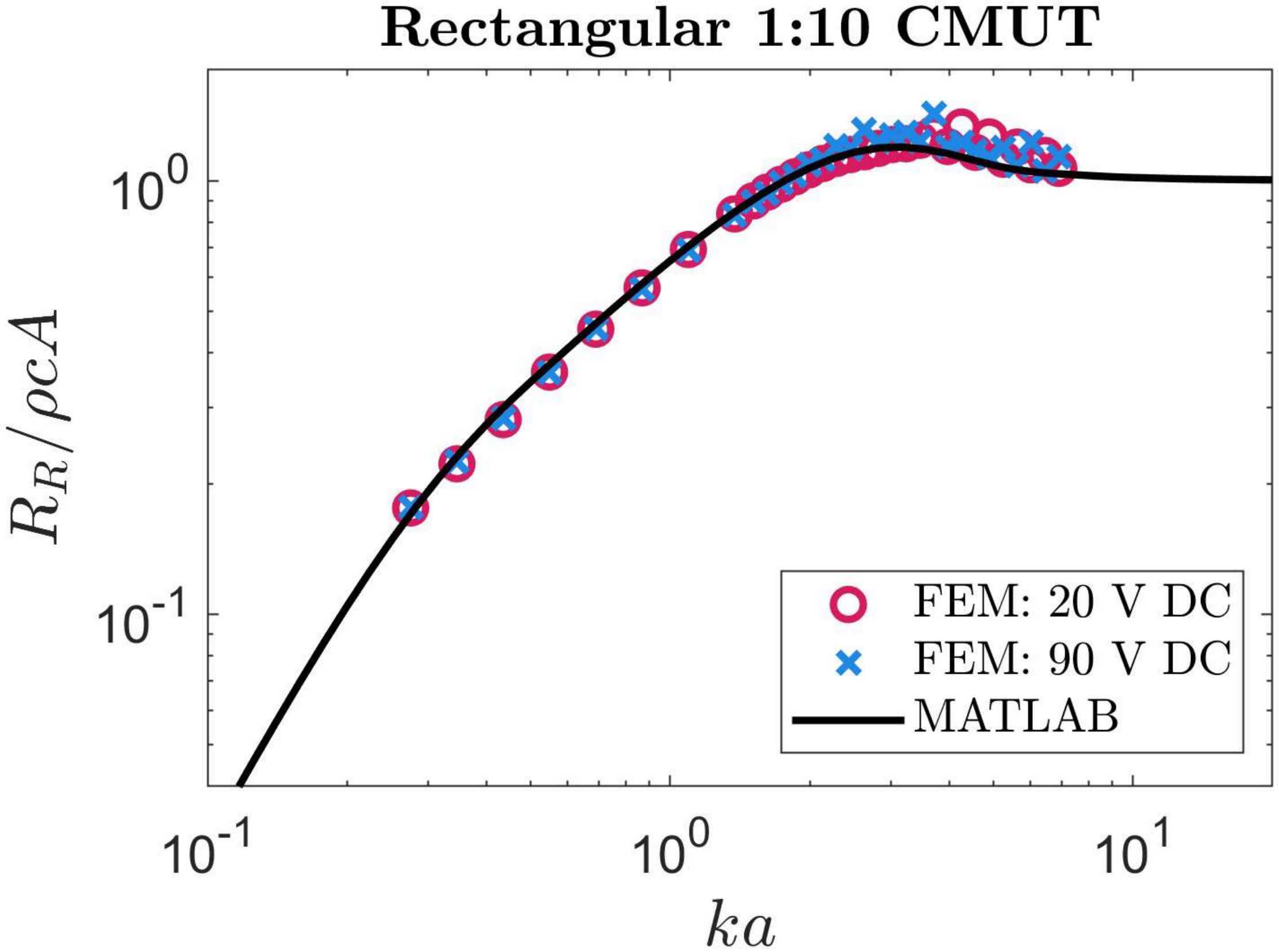}
\includegraphics[width=8cm]{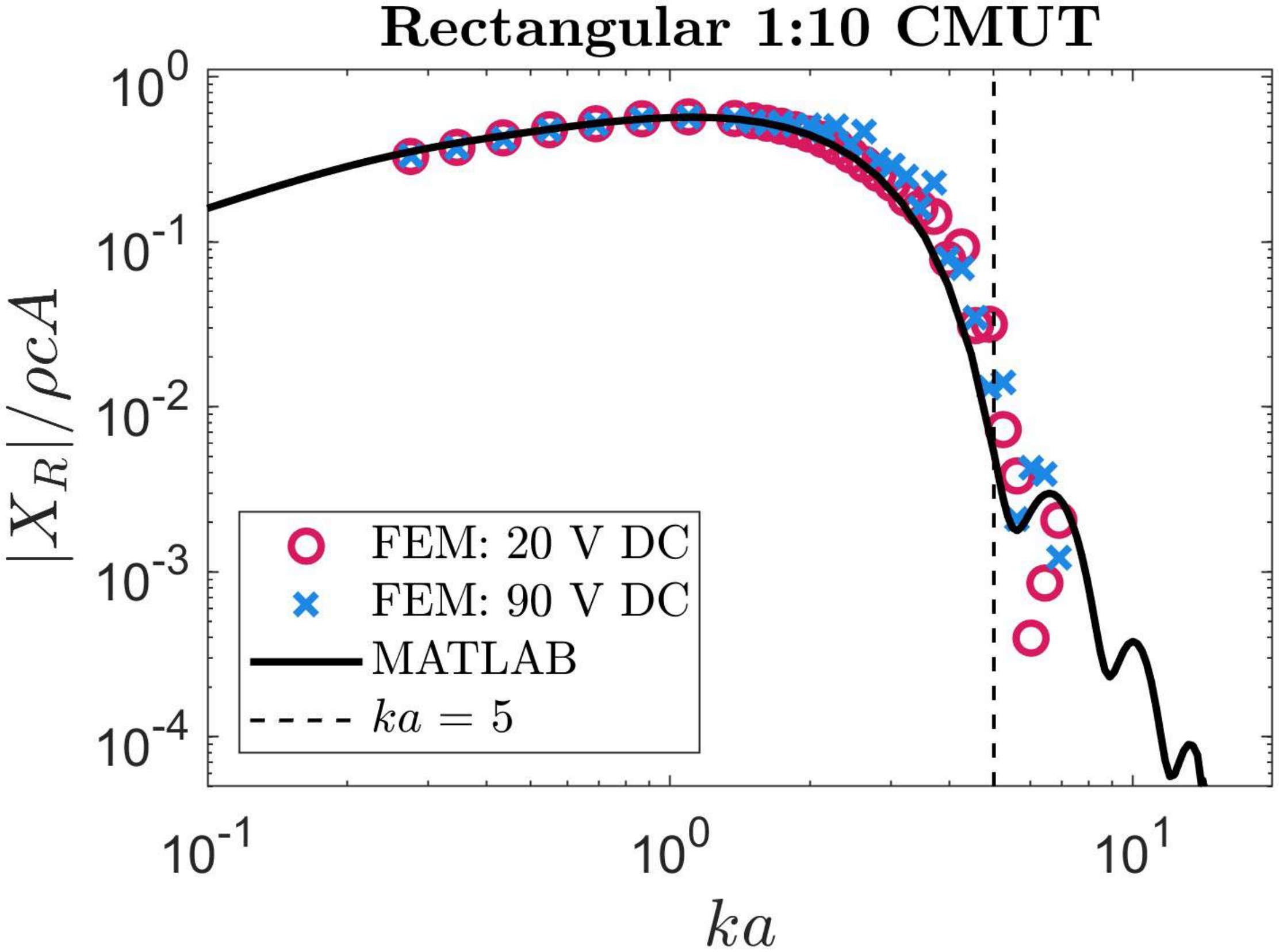}
\end{adjustwidth}
\caption{The radiation impedance of a rectangular 1:10 CMUT extracted from FEM compared to our method. \label{r10_ZR}}
\end{figure}

\begin{figure}[H]
\begin{adjustwidth}{-\extralength}{0cm}
\centering
\includegraphics[width=8cm]{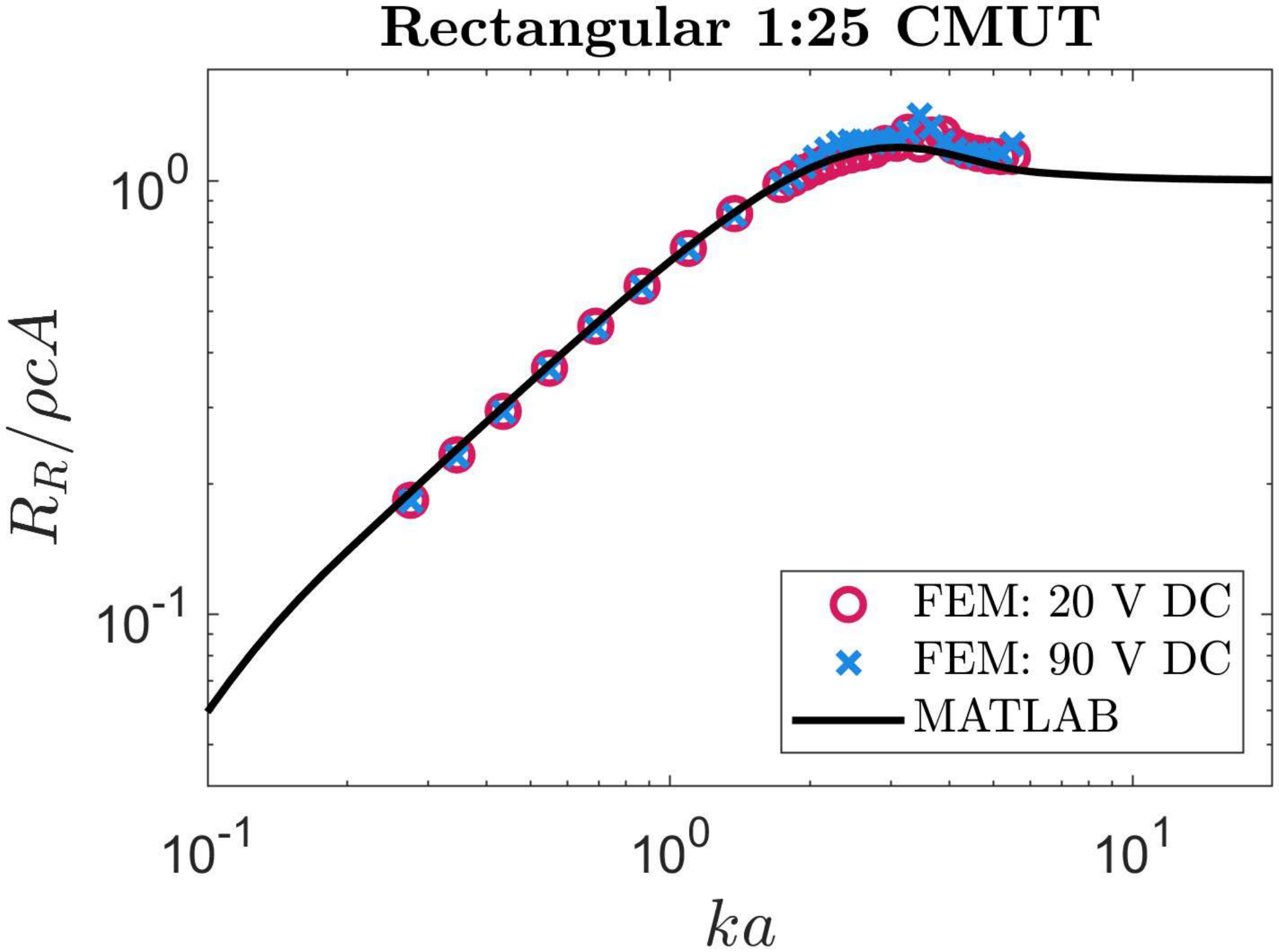}
\includegraphics[width=8cm]{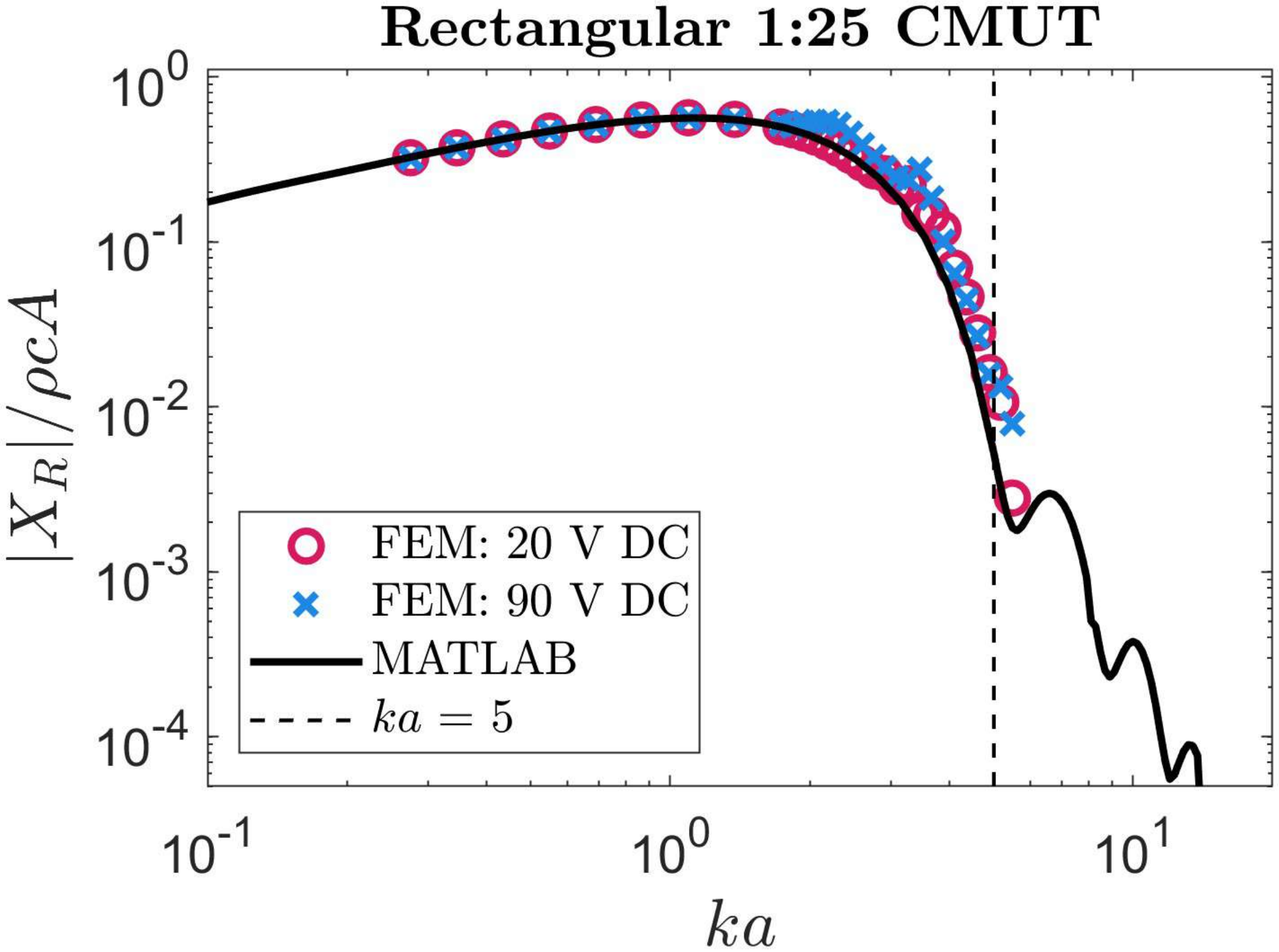}
\end{adjustwidth}
\caption{The radiation impedance of a rectangular 1:25 CMUT extracted from FEM compared to our method. \label{r25_ZR}}
\end{figure}

Our analytical expressions for a rectangular membrane were examined using the same approach. The FEM simulations used in Section \ref{FEMVal} to validate our theoretical velocity profiles were also able to produce radiation impedance values. These simulated results were compared to numerical evaluations of the integrals in \eqref{impR} and \eqref{impI} with a MATLAB script. Our theoretical predictions are compared to FEM results in Figures~\ref{sq_ZR}-\ref{r25_ZR}.

Similar to the circular case, there is reasonable agreement between FEM and theory across all frequencies for the radiation resistance. These agreements are quantified in Table~\ref{tab_MAPE}. For the reactance, we again observed a threshold above which our analytical approach was inaccurate, likely due to neglected higher-order modes. In this case, the threshold was $ka=5$. Differences between our approach and FEM for radiation reactance are also quantified in Table~\ref{tab_MAPE}. Notably, the reactance became less accurate at aspect ratios where our velocity profile was less accurate (as reported in Table~\ref{tab_ARE}). A noteworthy observation is that on average, this approach could compute the radiation impedance of each $ka$ faster than FEM by 4 orders of magnitude.

\subsection{Long Rectangular CMUT 1D Approximation}
\begin{figure}[H]
\begin{adjustwidth}{-\extralength}{0cm}
\centering
\includegraphics[width=8cm]{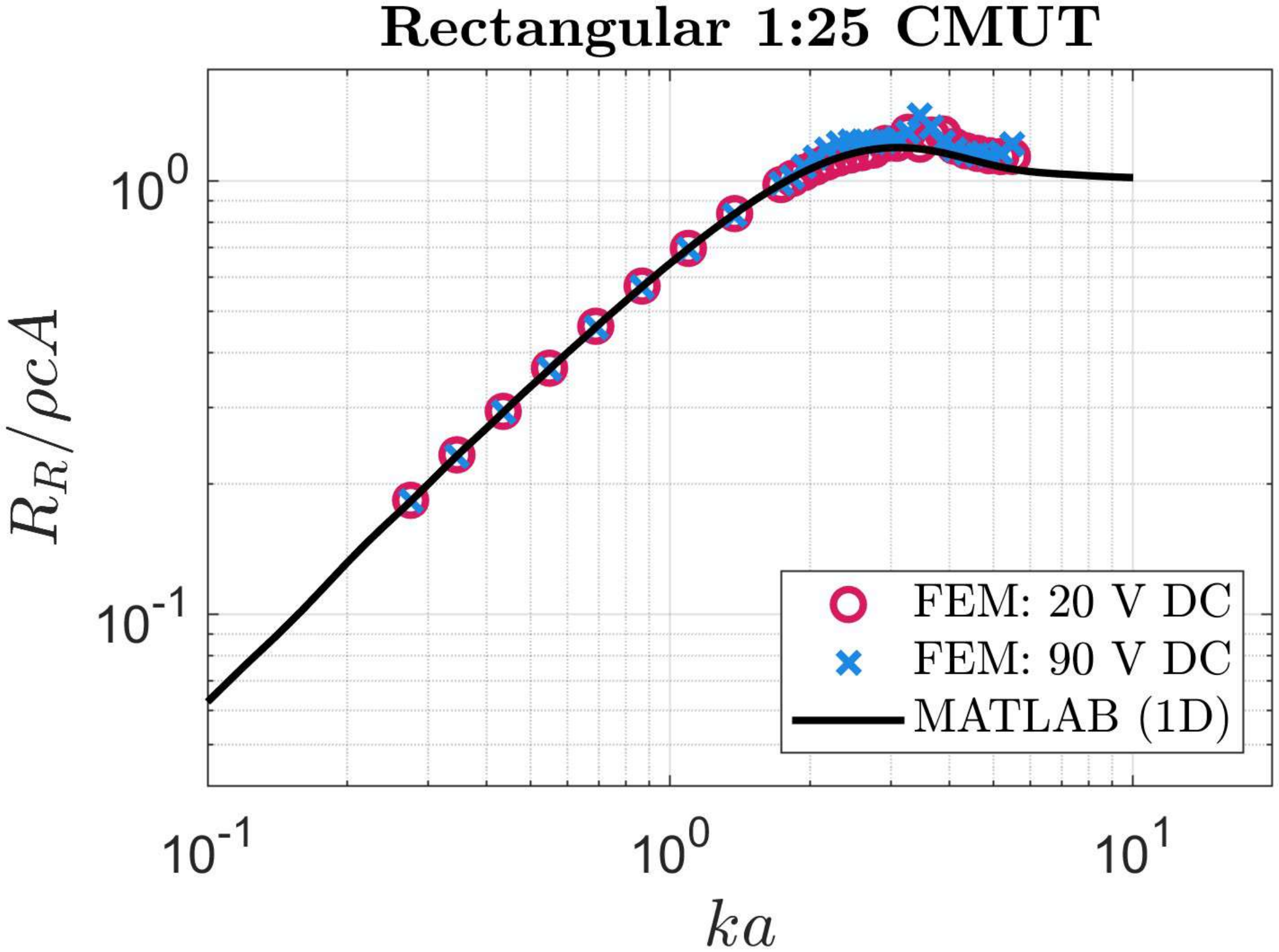}
\includegraphics[width=8cm]{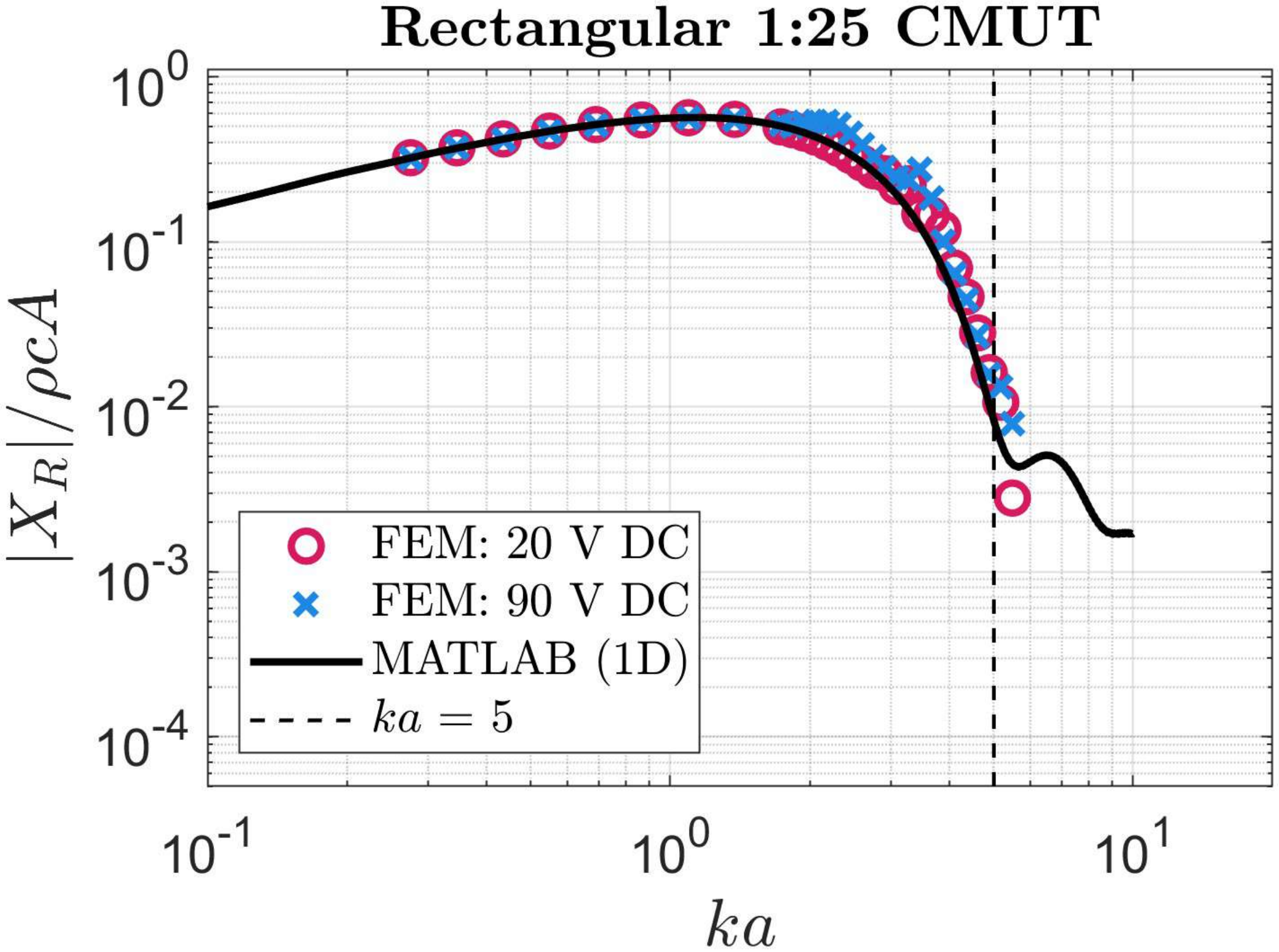}
\end{adjustwidth}
\caption{The radiation impedance of a rectangular 1:25 CMUT extracted from FEM compared to our method based on the 1D approximate velocity profile detailed in Appendix~\ref{1dapprox}. \label{r25_1d_ZR}}
\end{figure}

Using the 1D approximation, the expressions of \eqref{impR1d} and \eqref{impI1d} were derived in the Appendix~\ref{1dapprox}. Similar to our approach for regular rectangular membranes, the radiation impedance of a high aspect ratio rectangular membrane could be estimated using these expressions. The resulting radiation impedance obtained by numerical integration with MATLAB is compared with the values obtained from FEM for the 1:25 rectangular case in Figure~\ref{r25_1d_ZR}. 

The radiation impedance obtained by this approximation resulted in slightly better agreement with FEM compared to the original approximation which was based on \eqref{v(x,y)}. Similar to the previous cases, the threshold of $ka=5$ was used for comparing the calculated reactance with FEM. The discrepancies between the radiation impedance based on this 1D approximation and the FEM values are quantified in Table~\ref{tab_MAPE}.

\begin{table}[H] 
\caption{This table contains the mean absolute error (MAE) of the radiation impedance obtained using our approach compared to results obtained from FEM. The FEM simulations were performed with biases of $\sim 20\%$ and $\sim 90\%$ of the collapse voltage. The errors for the imaginary (reactive) part are calculated for $ka$ values below the threshold specified in Figures~\ref{circ_ZR}-\ref{r25_1d_ZR}. \label{tab_MAPE}}
\begin{adjustwidth}{-\extralength}{0cm}
\newcolumntype{C}{>{\centering\arraybackslash}X}
\begin{tabularx}{\fulllength}{CC|CC|CC}
\toprule
\multicolumn{2}{c|}{} & \multicolumn{2}{c|}{$MAE$ at low DC bias ($\times 10^{-2}$)} & \multicolumn{2}{c}{$MAE$ at high DC bias ($\times 10^{-2}$)} \\
\multicolumn{2}{c|}{Plate}	& $\frac{R_R}{\rho c A}$ & $\frac{X_R}{\rho c A}$	& $\frac{R_R}{\rho c A}$ & $\frac{X_R}{\rho c A}$\\
\midrule
  \multicolumn{2}{c|}{Circular \textsuperscript{1}} & 0.33 & 0.37 & 0.69 & 0.74 \\
  \multicolumn{2}{c|}{Rectangular with $b/a=1$} & 0.89 & 1.12 & 0.38 & 0.23 \\
  \multicolumn{2}{c|}{Rectangular with $b/a=4$} & 3.53 & 1.60 & 3.57 & 3.14 \\
  \multicolumn{2}{c|}{Rectangular with $b/a=10$} & 3.78 & 1.40 & 5.13 & 4.25 \\
  \multicolumn{2}{c|}{Rectangular with $b/a=25$} & 2.96 & 1.34 & 5.30 & 4.94 \\
  \multicolumn{2}{c|}{Rectangular with $b/a=25$ (1D approximation)} & 2.46 & 1.25 & 4.67 & 4.67 \\
\bottomrule
\end{tabularx}
\end{adjustwidth}
\noindent{\footnotesize{\textsuperscript{1} Comparison made with the expression derived in \cite{martin_greenspan}.}}
\end{table}

The improvement in calculation time with our numerical approach compared to FEM can be highlighted despite some compromises in accuracy. Due to hardware limitations, our COMSOL simulations had to be completed in multiple steps. First, a coarse frequency sweep was performed, followed by a second simulation with higher frequency resolution and finer meshing to ensure the high-frequency behavior was accurately captured. 

To put the speed advantage of our method into perspective, we calculated the average run-time for the computation of radiation impedance per frequency ($ka$ value). For the 1:25 rectangular case, the FEM simulations took approximately \qty{1050}{s} to produce each of the points plotted in Figures~\ref{r25_ZR} or \ref{r25_1d_ZR}. In contrast, our original approach resulted in \qty{0.13}{s} average run-time for each $ka$ plotted in these figures. The 1D approach was slightly slower at \qty{44}{s} per $ka$. It should be highlighted that even the 1D approximation approach as the slower method has a speed advantage of about 2 orders of magnitude compared to FEM.

\section{Discussion} \label{discuss}

This work is among the first papers to investigate the radiation impedance of clamped rectangular plates with larger aspect ratios. These high-aspect-ratio rectangular radiators are of particular interest due to the promising results shown with long single-membrane rectangular CMUTs \cite{ericdew}. Such rectangular CMUTs would benefit heavily from analytical models to optimize their designs and better understand their operation. Thus, we believe our method will address a previously unmet need for radiation impedance models applicable to those designs.

In terms of radiation resistance, agreement was close between our analytical expression and FEM simulations across all frequencies for a variety of geometries. Our model had a mean absolute error of $0.053 \rho c A$ or better for all designs when compared with simulated CMUT membranes at high and low bias voltages. For radiation reactance, results at high frequency ($ka>5$) did not match our simulation due to higher order modes not being accounted for in our velocity profile. However, our results below $ka=5$ had reasonable agreement with FEM. This reduced frequency range is not a major limitation, as most CMUTs are designed based on their fundamental mode which occurs below this threshold \mbox{\cite{koymen, improved_koymen}}. Other works have also reported challenges modeling radiation impedance at high frequencies and limited their scope to a similar range of frequencies \mbox{\cite{Dingguo2009, gomperts1977sound}}. We used a similar procedure to verify the well-known expressions for the radiation impedance of a circular clamped radiator and observed similar deviations between analytical results and FEM at high frequency.

At higher (length:width) aspect ratios, our approximate velocity profile became less accurate and thus impacted the accuracy of the radiation reactance. For these very large aspect ratios, we further calculated results using a “1D” approximate velocity profile (analogous to the case of a plate with the long edges clamped and short edges free). This “1D” approximate velocity profile yielded more accurate results for a plate with an aspect ratio of 25.

It is likely that even more accurate results could be obtained for high aspect ratios with further attention to approximating the velocity profile at these dimensions. This could be performed in future work, using our method as a starting point and simply substituting a more precise velocity profile. It would also be valuable to consider higher order modes, as many works have demonstrated that such modes occur closer in frequency to the fundamental mode at large aspect ratios \cite{Wong2008, Warburton1954}. 

An additional challenge with obtaining an accurate velocity profile for high aspect ratio rectangular membranes is modeling the impacts of stress. Diaphragms are generally categorized as either membranes, thin plates, or thick plates, depending on the dimensions of the layer. Circular structures are categorized based on the radius:thickness ratio, with a ratio of < 10 corresponding to a thick plate, a ratio of > 80 corresponding to a membrane, and a value between those boundaries corresponding to a thin plate \cite{Ventsel2001txtbook}. In rectangular devices, the challenge is that one dimension may be categorized as a membrane, while the other may be more accurately described as a thin or thick plate. Rather than selecting dimensions to fit within a certain regime, our simulated dimensions were selected to be similar to previously demonstrated rectangular CMUT devices. In this case, the width:thickness ratio was 31, while the length:thickness ratio varied between 31 and 600.  As our velocity profile was intended to describe thin plates, it is unsurprising that it accurately described velocity along the short dimension (thin plate), and lost accuracy as the long dimension entered into the membrane category. 
Other works have derived expressions to describe membrane deflection, which may be suitable to describe the velocity profile along the long dimension in future work \cite{zheng2015novel}.

%%%%%%%%%%%%%%%%%%%%%%%%%%%%%%%%%%%%%%%%%%
\section{Conclusions}
This study advances the modeling of capacitive micromachined ultrasound transducers (CMUTs) with rectangular membranes by introducing an efficient method to approximate their acoustic radiation impedance. Using a polynomial shape model for the velocity profile, this approach offers a rapid means of computing radiation impedance, which is validated against finite element method (FEM) simulations across various frequencies and aspect ratios. Our proposed method can output the radiation resistance with good accuracy and provides reasonable predictions for the radiation reactance at frequencies below $ka=5$. Although accuracy slightly decreases at high aspect ratios, there are limited models available for these dimensions; our 1D approximate velocity profile may address this unmet need and enable radiation impedance calculations with acceptable accuracy. These results could be used to accelerate the design of rectangular CMUTs and optimize their performance, potentially enabling the development of more advanced ultrasound transducer technologies.

%%%%%%%%%%%%%%%%%%%%%%%%%%%%%%%%%%%%%%%%%%
\vspace{6pt} 

%%%%%%%%%%%%%%%%%%%%%%%%%%%%%%%%%%%%%%%%%%
%% optional
\supplementary{The MATLAB script developed for the computation of radiation impedance is available online at \cite{github_repo}.} % \linksupplementary{s1}

%%%%%%%%%%%%%%%%%%%%%%%%%%%%%%%%%%%%%%%%%%
\authorcontributions{Analytical methods, S.K.; Numerical methods, E.D. and S.K.; Simulations, S.K. and M.S.; Writing, E.D. and S.K.; Supervision, R.Z
All authors have read and agreed to the published version of the manuscript.}

\funding{
We gratefully acknowledge funding from Alberta Innovates (AB Innovat CASBE 212200391), NSERC (AACASBE 567531 - 21), and the National Institutes of Health (1R21HL161626-01 and EITTSCA R21EYO33078).
}

\institutionalreview{Not applicable.}

\informedconsent{Not applicable.}

\dataavailability{The original contributions presented in the study are included in the supplementary material, further inquiries can be directed to the corresponding authors.} 

\acknowledgments{We are grateful to CMC
Microsystems for access to the CAD tools used in this work.}

\conflictsofinterest{The authors declare no conflicts of interest.}

%%%%%%%%%%%%%%%%%%%%%%%%%%%%%%%%%%%%%%%%%%
%% Optional
\appendixtitles{yes} 
\appendixstart
\appendix
\section[\appendixname~\thesection]{Change of Variables for Integration} \label{changeVars}
This section describes the change of variables utilized in deriving \eqref{impR} and \eqref{impI} from \eqref{power2} which is based on \cite{TimMellow}. Starting with the expression for the wavevector magnitude in Cartesian coordinates:
\begin{linenomath}
    \begin{equation} \label{K}
        k=\sqrt{k_x^2 + k_y^2 + k_z^2},
    \end{equation}
\end{linenomath}

Let us define $t$ as:
\begin{linenomath}
    \begin{equation}
        t = \frac{\sqrt{k_x^2 + k_y^2}}{k} 
    \end{equation}
\end{linenomath}

Then, the x and y components of $\textbf{k}$ can be obtained as:
\begin{linenomath}
    \begin{equation}
k_x=kt\cos\phi
\end{equation}
\end{linenomath}
\begin{linenomath}
    \begin{equation}
k_y=kt\sin\phi        
    \end{equation}
\end{linenomath}

Hence, the differential for $k_x$ and $k_y$ can be rewritten as:
\begin{linenomath}
    \begin{equation}
\,dk_x\,dk_y = k^2 t\,dt\,d\phi        
    \end{equation}
\end{linenomath}

Consequently, $k_z$ can be expressed differently using its definition in \eqref{k_z} and $t$ as:

\begin{linenomath}
    \begin{equation}
        k_z = \Biggl\{ \begin{array}{cc} k\sqrt{1-t^2} & 0<t\le 1 \\ 
        -jk\sqrt{t^2-1} & 1<t<\infty \\ \end{array}
    \end{equation}
\end{linenomath}

Using the derived expressions for $k_x$, $k_y$, and $k_z$, expression of \eqref{impR} and \eqref{impI} can be derived from \eqref{power2}.

%%%%%%%%%%%%%%%%%%%%%%%%%%%

\section[\appendixname~\thesection]{1D Velocity Profile Approximation for Long Rectangular Plates} \label{1dapprox}

The 1-dimensional approximation to the velocity profile of a long rectangular membrane is detailed below.

\begin{linenomath}
    \begin{equation} \label{v1d}
        v(x,y) = v_0 \left( 1 - \left(\frac{x}{a}\right)^2 \right)^2 \Pi(\frac{y}{2b}),
    \end{equation}
\end{linenomath}
where $a$ is the half-width of the rectangle, and $b$ is its half-length. The function $\Pi(\frac{y}{2b})$ is the rectangular functions, which is equal to 1 where $|y|<b$ and zero everywhere else. The derivations in the following paragraphs were performed with Wolfram Mathematica.

Following the procedure outlined in Section~\ref{derivation}, the total emitted power can be obtained through the derivation of the pressure phasor. This intermediate step of deriving the pressure is omitted for the sake of brevity. The x-dependent portion of the 1D velocity profile is identical to the x-dependent half of the original shape function \eqref{v(x,y)}. This similarity leads to the appearance of $S(k_xa)$ inside the expression for power as stated below. The function $S(k_xa)$ is defined in \eqref{Sx}. The y-dependent portion of the power integral is proportional to the Fourier Transform of $\Pi(\frac{y}{2b})$, corresponding to multiples of $\text{Sinc}(k_yb)$.

\begin{linenomath}
    \begin{equation} \label{Ptot1d}
        P_{tot} = \frac{\rho c k v_0^2 a^2 b^2}{\pi^2} \iint_{-\infty}^{\infty} \ S^2(k_xa)\ \text{Sinc}^2(k_yb) \frac{\,dk_x\,dk_y}{k_z}
    \end{equation}
\end{linenomath}

After applying the same change of variables (detailed in Appendix~\ref{changeVars}) and calculating RMS velocity based on \eqref{v1d} and \eqref{VRMS}, we can generate the expressions for the real and imaginary part of radiation impedance according to \eqref{RI_def}.
%"the calculation of RMS velocity" -> calculating RMS velocity
%After applying the same change of variables (detailed in Appendix~\ref{changeVars}) and the calculation of RMS velocity based on \eqref{v1d} and \eqref{VRMS}, we can generate the expressions for the real and imaginary part of radiation impedance according to \eqref{RI_def}.

\begin{adjustwidth}{-\extralength}{0cm}
    \begin{equation}\label{impR1d}
    R_R = \frac{\rho c k^2 a^2 b^2}{\pi^2} \ \frac{315}{128} \\ \int_{0}^{2\pi} \int_{0}^{1} S^2(kta\cos\phi)\ \text{Sinc}^2(ktb\sin\phi) \frac{t\,dt\,d\phi}{\sqrt{1-t^2}}
\end{equation}
\end{adjustwidth}

\begin{adjustwidth}{-\extralength}{0cm}
    \begin{equation}\label{impI1d}
    X_R = - \frac{\rho c k^2 a^2 b^2}{\pi^2} \ \frac{315}{128} \\ \int_{0}^{2\pi} \int_{1}^{\infty} S^2(kta\cos\phi)\ \text{Sinc}^2(ktb\sin\phi) \frac{t\,dt\,d\phi}{\sqrt{t^2-1}}
    \end{equation}
\end{adjustwidth}

Similar to the investigated 2D case, these integrations can be computed numerically, although further issues with convergence were encountered. We speculate that these difficulties arise from $Sinc(x)$ approaching zero more slowly than $S(x)$. These challenges were resolved in the provided MATLAB script by evaluating the limits as $x \rightarrow \infty$ \cite{github_repo}. Evaluating these integrals yields the radiation impedance results plotted in Figure~\ref{r25_1d_ZR}. Agreement between these values and our FEM results is quantified in Table~\ref{tab_MAPE}.

%%%%%%%%%%%%%%%%%%%%%%%%%%%%%%%%%%%%%%%%%%
\begin{adjustwidth}{-\extralength}{0cm}

\reftitle{References}

\bibliography{external_bib}

\end{adjustwidth}
\end{document}